\documentclass[12pt,a4paper]{article}

\usepackage{amssymb}
\usepackage[dvips]{color}
\usepackage[centerlast,footnotesize]{caption2}
\usepackage{float}

\usepackage{multirow}
\usepackage{booktabs}
\usepackage{latexsym}
\usepackage{amsmath}
\usepackage{amsbsy}
\usepackage{epsfig}
\usepackage{graphicx}
\usepackage{latexsym}
\usepackage{euscript}
\usepackage{amsfonts}
\usepackage{mathtools}
\usepackage{braket}
\usepackage{simplewick}
\usepackage{amsmath}
\usepackage{amssymb}
\usepackage{dsfont}
\usepackage{slashed}
\usepackage{verbatim}

\newcommand{\lapprox}{\raisebox{-0.5ex}{$\
\stackrel{\textstyle<}{\textstyle\sim}\ $}}
\newcommand{\gapprox}{\raisebox{-0.5ex}{$\
\stackrel{\textstyle>}{\textstyle\sim}\ $}}

\usepackage{graphicx}
\newcommand{\One}{1\kern-4.5pt1}

\newcommand{\be}{\begin{equation}}
\newcommand{\ee}{\end{equation}}

\def\lesim{${\lower 2pt\hbox{$\scriptstyle
<$}\atop\raise 4pt\hbox{$\scriptstyle\sim$}}$} 
\def\grsim{${\lower2pt\hbox{$\scriptstyle >$} \atop\raise4pt\hbox 
{$\scriptstyle\sim$}}$} 

\begin{document}
\begin{center}
\begin{flushright}
September 2020
\end{flushright}
\vskip 10mm
{\LARGE
Critical Behaviour in the Single Flavor\\ Thirring Model in 2+1$d$
}
\vskip 0.3 cm
{\bf Simon Hands$^a$, Michele Mesiti$^b$, and Jude Worthy$^a$}
\vskip 0.3 cm
{\em $^a$ Department of Physics, College of Science, Swansea University,\\
Singleton Park, Swansea SA2 8PP, United Kingdom.}
\vskip 0.3 cm
{\em $^b$Swansea Academy for Advanced Computing, Swansea University,\\
Bay Campus, Swansea SA1 8EN, United Kingdom.}
\end{center}

\vskip 1.3 cm
\noindent
{\bf Abstract:} 
Results of a lattice field theory simulation of 
the single-flavor Thirring model in 2+1 spacetime dimensions are presented.
The lattice model is formulated using domain wall fermions as a means to
recover the correct U(2) symmetries of the continuum model in the limit where 
wall separation $L_s\to\infty$.
Simulations on $12^3,16^3\times L_s$, varying self-interaction strength $g^2$
and bare mass $m$, are
performed with $L_s=8,\ldots,48$,  and the results for the bilinear condensate
$\langle\bar\psi\psi\rangle$ fitted to a model equation of state assuming a 
U(2)$\to$U(1)$\otimes$U(1)
symmetry-breaking phase transition at a critical $g_c^2$. 
First estimates for $g_c^{-2}a$ and critical exponents
are presented, showing small but significant departures from mean-field values. 
The results confirm that
a symmetry-breaking transition does exist and therefore the critical number of
flavors for the Thirring model $N_c>1$. 
Results for both condensate and associated susceptibility are also obtained in
the broken phase on
$16^3\times48$, suggesting that here the $L_s\to\infty$ extrapolation is not yet
under control. We also present results obtained with the associated
2+1$d$ truncated overlap operator $D_{OL}$ demonstrating exponential localisation, a
necessary condition for the recovery of U(2) global symmetry, but that 
recovery of the Ginsparg-Wilson condition as $L_s\to\infty$ is extremely slow in
the broken phase. 

\vspace{0.5cm}


\newpage

\section{Introduction}
\label{sec:intro}
The Thirring model is a quantum field theory of relativistic fermions
interacting via a contact between conserved currents. In this paper we will 
examine the model in 2+1$d$ spacetime dimensions, and therefore further
specify the use of reducible (ie. four-component) spinor fields, permitting the
formulation of a parity-invariant mass term. The Lagrangian density reads
\begin{equation}
{\cal
L}=\bar\psi_i(\partial\!\!\!/\,+m)\psi_i+{g^2\over{2N}}(\bar\psi_i\gamma_\mu\psi_i)^2
\label{eq:contThir}
\end{equation} 
where $\mu=0,1,2$ and a sum over $i$ indexing $N$ fermion species is implied.
For a reducible representation of the Dirac algebra, there are two independent
matrices $\gamma_3$ and $\gamma_5$ which anticommute with the kinetic term in
(\ref{eq:contThir}), which results in a U(2$N$) global symmetry generated by the
set $\{\One,\gamma_3,\gamma_5,i\gamma_3\gamma_5\}$. For $m\not=0$ this breaks as
U$(2N)\to$U$(N)\otimes$U($N$).  The symmetry can also break spontaneously
through generation of a bilinear condensate $\langle\bar\psi\psi\rangle\not=0$; 
while there are clear analogies with ``chiral'' symmetry
breaking, this nomenclature should be eschewed in odd spacetime dimension.

Spontaneous U($2N$) symmetry breaking is believed to be theoretically possible
for sufficiently large self-coupling $g^2$ and sufficiently small $N$. Previous
investigations of this question have used truncated Schwinger-Dyson
equations~\cite{Gomes:1990ed,Itoh:1994cr,Sugiura:1996xk},
and Functional Renormalisation
Group~\cite{Gies:2010st,Janssen:2012pq,Gehring:2015vja}. 
The prototype scenario has a symmetry breaking transition at
$g^2_c(N)$, where $g^2_c$ is a increasing function of $N$. 
A UV-stable renormalisation group fixed point can be defined as $g^2\to g_c^2$, 
where we identify a {\em Quantum Critical Point\/} (QCP) 
such that there exists an interacting continuum
field theory solely specified by the field content, dimensionality and pattern of
symmetry breaking. 
The critical flavor number 
$N_c$ required for symmetry breaking defined by
$g^2_c(N)\vert_{N=N_c}\to\infty$, which in principle need not be integer,
is an important property of the model. Since
symmetry breaking is not accessible via expansion in any small parameter, the
identification of $N_c$ for the Thirring model
is an important and challenging problem in non-perturbative quantum field theory. 

The model also provides a natural arena for the application of lattice field
theory methods,
and may well be the simplest fermion field theory {\em requiring\/} a
computational solution. It turns out the choice of fermion discretisation has
undue influence. Many early
studies~\cite{DelDebbio:1997dv,Kim:1996xza,DelDebbio:1999he,Hands:1999id,Christofi:2007ye,
Chandrasekharan:2011mn}
used staggered fermions,
which naturally support a symmetry breaking
U(${N\over2})\otimes$U(${N\over2})\to$U(${N\over2}$)~\cite{Burden:1986by}.
The results of these studies are broadly consistent; the most wide-ranging of
them,
exploiting plausible assumptions about the $g^2\to\infty$ limit on a
lattice, found $N_c=6.6(1)$~\cite{Christofi:2007ye}. The critical exponents
found for $N<N_c$ appear rather sensitive to $N$, suggesting the existence of a
rich family of distinct QCPs. 

There are reasons to question whether this prediction for $N_c$ is correct;
issues arising from a lattice perspective are reviewed
in \cite{Hands:2018vrd}. Here we present a non-rigorous plausibility argument for
caution based on symmetry. The Thirring model in 2+1$d$ may be studied
analytically in the limit of large $N$, and the mass $M_V$ of a vector $f\bar f$
bound state interpolated by the current density $\bar\psi\gamma_\mu\psi$ is predicted in this
limit to be~\cite{Hands:1994kb}
\begin{equation}
{M_V\over m}=\sqrt{{6\pi}\over{mg^2}};\;\;\;\lim_{g^2\to\infty}{M_V\over
m}=0.
\end{equation}
Hence in the strong-coupling limit the Thirring model is a theory of conserved
currents interacting via exchange of a massless spin-1 boson. Asymptotically the
boson propagator switches from the canonical $D_{\mu\nu}(k)\propto k^{-2}$ to
$\propto k^{-1}$;
this UV behaviour is exactly that predicted for the photon of QED$_3$ in the IR
limit. Hence the Thirring QCP, if it exists, could well be identical with the IR
fixed point of QED$_3$, and the critical $N_c$ for both theories should then
coincide. The parallels between the Thirring model and abelian gauge theory
are more apparent still once an auxiliary vector field $A_\mu$ is introduced,
as reviewed in Sec.~\ref{sec:formulation} below and more generally throughout
the rest of the paper. 

Now, there is an old argument~\cite{Appelquist:1999hr} for estimating $N_c$ in
QED$_3$, based on the conjecture $f_{IR}\leq f_{UV}$, where 
\begin{equation}
f_{IR} = -{90\over\pi^2} \lim_{T\to0}{{\cal F}\over T^4};\;\;\
f_{UV} = -{90\over\pi^2} \lim_{T\to\infty}{{\cal F}\over T^4},
\end{equation}
and ${\cal F}$ is the thermodynamic free energy density.
For asymptotically-free theories such as QED$_3$ $f_{UV}$ is related to the
count of non-interacting constituents:
\begin{equation}
f_{UV}={3\over4}\times4N+1.
\end{equation}
Here ${3\over4}$ is the appropriate factor for Fermi-Dirac statistics in 2+1$d$,
4 is the number of spinor components per flavor and 1 counts the single
physical polarisation state of the photon, which remains
unconfined in QED$_3$.
For a phase with spontaneously-broken symmetry the number of weakly-interacting
degrees of freedom is the Goldstone count plus the photon. Hence
\begin{equation}
f_{IR}=\begin{cases}2N^2+1&{\rm U}(2N)\to {\rm U}(N)\otimes{\rm  U}(N);\\
{N^2\over4}+1&{\rm U}({N\over2})\otimes{\rm U}({N\over2})\to{\rm U}({N\over2}).\end{cases}
\end{equation}
For QED$_3$, and by extension the continuum Thirring model, the conjecture therefore
predicts $N_c\leq{3\over2}$, whereas for the symmetry breaking dictated
by the staggered formulation the equivalent bound is the much less stringent
$N_c\leq12$. This disparity is a strong motivation for exploring lattice
fermions with the correct global symmetry.

Recently this programme has been developed in two distinct directions. Lattice
fermions employing the SLAC derivative, which is 
non-local but manifestly U($2N$)-symmetric, have been used to investigate the
model in \cite{Wellegehausen:2017goy,Lenz:2019qwu}. Since the Thirring model is
not a gauge theory, the long-standing objections \cite{Karsten:1979wh} 
to the SLAC approach associated with lack of localisation of the fermion-gauge
vertex do not
apply. No evidence has been
found for spontaneous symmetry breaking for any integer value of $N$ (ie. any
unitary local version of the model); an estimate $N_c=0.80(4)<1$ was reported
in \cite{Lenz:2019qwu}. Meanwhile, the Thirring model formulated 
with domain wall fermions (DWF) has
been investigated by us in \cite{Hands:2016foa,Hands:2018vrd}. DWF employ a
local lattice derivative at the cost of introducing a fictitious extra
dimension. In this case the
U($2N$) symmetry is not manifest but hopefully recovered in a controlled way 
in the limit that the separation
$L_s$ between domain walls located at either end of this ``third'' dimension is
made large. An aspect worth highlighting is that in the most promising approach 
the fermions interact with 
the auxiliary field $A_\mu$ throughout the bulk, ie for $0<x_3<L_s$, very
similar to the way gauge theories are formulated with DWF~\cite{Shamir:1993zy}.

We have found that the $L_s\to\infty$ limit becomes particularly
challenging precisely in the strong-coupling regime where symmetry breaking is
anticipated. In particular, Ref.~\cite{Hands:2018vrd} presented results for 
$N=0,1,2$ on 
$12^3\times L_s$ (for $N>0$) with $L_s$ ranging from 8 to 40. The results for
$N=0$ and $N=2$ are fairly clear-cut:  symmetry-breaking is observed in the
former case and not in the latter. For $N=1$ there are qualitative indications
of a change in the system's behaviour at the strongest coupling explored
($g^{-2}a=0.3$, where the lattice spacing $a$ defines the scale). In particular 
the bilinear condensate $\langle\bar\psi\psi(m)\rangle$ displays significant
$L_s$-dependence in this regime, and the resulting signal estimated in the
$L_s\to\infty$ limit is significantly greater than that from weaker couplings;
there is also a marked departure from linear $m$-dependence. 
The conclusion reached in \cite{Hands:2018vrd} is that $0<N_c<2$ with ``strong
evidence'' for $N_c>1$.

Settling the issue is hard for a couple of reasons. One is that the RHMC
simulation algorithm required for $N=1$
is numerically more
demanding~\cite{Hands:2018vrd}. Another is that in a finite volume the
bilinear condensate vanishes identically for $m=0$ so that neither order parameter
nor associated susceptibility are directly accessible in the U($2N$)-symmetric
limit. In the current paper we apply improved code and substantially enhanced
computing resources to both issues, by exploring the model on both $12^3\times L_s$
and $16^3\times L_s$, with $L_s$ as large as 48, with much finer resolution along the
coupling axis, particularly in the suspected critical region. The resulting order
parameter data $\langle\bar\psi\psi(g^2,m)\rangle$ are obtained with sufficient
statistical precision to permit fits to a renormalisation-group inspired
equation of state (EoS) applicable away from $m=0$, yielding estimates for both
critical coupling $g^2$ and accompanying critical exponents $\beta_m$ and
$\delta$ defined in (\ref{eq:crit_exponents}) below. A similar strategy has 
been applied successfully in staggered fermion
studies~\cite{DelDebbio:1997dv,DelDebbio:1999he,Christofi:2007ye}.
As the methodology implies, the main results of the study are confirmation that
$N_c>1$ and a first tentative characterisation of the critical properties of the
QCP.

The remainder of the paper is organised as follows. Sec.~\ref{sec:formulation}
sets out the lattice formulation of the model using DWF and the auxiliary field
method, and reviews the simulation algorithm and principal observables.
Results are presented in Sec.~\ref{sec:results} in three subsections. 
Sec.~\ref{sec:Lsextrap} presents a systematic study of varying
$\beta\equiv g^{-2}a$, $m$,
$L_s$ and space-time volume, in the regime $0.3\leq\beta\leq0.52$.
Results for $\langle\bar\psi\psi\rangle$ 
are first extrapolated to $L_s\to\infty$, and then fitted to the
empirical EoS (\ref{eq:eos}) below. The picture that emerges 
appears under control; the $L_s$-extrapolation and EoS-fitting procedures
almost, but not quite, commute, yielding fairly stable estimates for
$\beta_c\approx0.28$
and estimates for the exponents $\beta_m,\delta$ distinct from those of mean
field theory. Finite volume effects appear remarkably small. 
However, all the data used in this analysis lie in the symmetric
phase. To address this Sec.~\ref{sec:Ls48} presents a complementary study at
fixed $L_s=48$, but this time exploring couplings as strong as $\beta=0.23$.
Both 
$\langle\bar\psi\psi\rangle$  and its associated susceptibility $\chi_\ell$ are studied. Here some issues
emerge; the EoS fit is not so successful at describing data in the strong-coupling phase,
and $\chi_\ell$, while having a peak as expected near criticality, exhibits a
non-standard scaling as $m$ is varied. We also present results for a residual
$\delta_h$ introduced in \cite{Hands:2015qha} to quantify recovery of U($2N$)
symmetry, and the bose action density $(2g^2)^{-2}\langle A^2_\mu\rangle$.
In Sec.~\ref{sec:locality} we present results for the locality, and recovery of
the Ginsparg-Wilson relation, 
of the truncated overlap operator (corresponding to finite
$L_s$)  in the critical region. Both are key
ingredients in demonstrating the existence of a local
U($2N$)-symmetric field theory at the QCP.
Our conclusions are discussed in 
Sec.~\ref{sec:discussion}.

\section{Formulation and Numerical Simulation}
\label{sec:formulation}

We use an auxiliary field formulation to represent the Thirring
interaction, which recasts the fermion action as a bilinear form while
preserving the global symmetries.
In continuum notation the Lagrangian density reads
\begin{equation}
{\cal L}^\prime=\bar\psi(\partial\!\!\!/\,+iA\!\!\!/\,+m)\psi+{1\over{2g^2}}A_\mu^2.
\end{equation}
On a lattice the vector auxiliary field $A_\mu$ is naturally formulated on a
link. Preserving fermion global symmetries while transcribing to a lattice is of
course a long-standing issue in lattice field theory.  Our approach uses domain wall
fermions (DWF), in which a fictitious third dimension is introduced so 
the fermions $\Psi(x,s)$ are defined in 2+1+1$d$:
\begin{equation}
S=\sum_{x,y}\sum_{s,s^\prime}\bar\Psi(x,s)[\delta_{s,s^\prime}D_{Wx,y}[A]
+\delta_{x,y}D_{3s,s^\prime}]\Psi(y,s^\prime)
+mS_m+{1\over{2g^2}}\sum_{x,\mu}A_\mu^2(x).
\label{eq:action}
\end{equation}
We use the M\"obius formulation, implying that $D_W[A=0], D_3$ are free Wilson
derivative operators, with $D_{3s,s^\prime}$ having open boundary conditions
implemented on the hopping terms, viz. 
$\delta_{s\mp1,s^\prime}(1-\delta_{s^\prime,1/L_s})$, at domain walls located at $s=1,L_s$.
The auxiliary field $A_\mu(x)$ is 3-static, taking the same value on every spacetime
slice along the 3rd direction. In previous work we have referred to this as the
{\em bulk\/} version of the model~\cite{Hands:2018vrd}. Our model employs a {\em non-compact} formulation
of the interaction in which each hopping term in $D_W$ carries a non-unitary
link factor
$(1\pm iA_\mu(x))$; in this way integration over $A$ generates solely 4-fermi
terms, and not higher point contact interactions. Further details are set out in 
\cite{Hands:2018vrd}.

In the large-$L_s$ limit the free kinetic operator approaches an
overlap operator $D_{OL}$ defined in 2+1$d$ and satisfying a generalisation of the
Ginsparg-Wilson relations~\cite{Ginsparg:1981bj,Hands:2015dyp}
\begin{equation}
\{\gamma_3,D_{OL}\}=2D_{OL}\gamma_3D_{OL};\;\;\;
\{\gamma_5,D_{OL}\}=2D_{OL}\gamma_5D_{OL};\;\;\;
[\gamma_3\gamma_5,D_{OL}]=0.
\label{eq:GW}
\end{equation}
For weakly-interacting fields, the RHS of the first two of 
these relations are formally $O(a)$
and thus
vanish in the continuum limit, recovering the desired U(2) global symmetry of
the continuum model. This only holds for the bulk formulation of the
Thirring model with DWF, and not the alternative ``surface'' formulation 
discussed in \cite{Hands:2016foa,Hands:2018vrd}. For strongly-interacting fields the
recovery of the GW relations (\ref{eq:GW}) and the locality of the corresponding
$D_{OL}$ will be discussed further in Sec.~\ref{sec:locality} below.

The bridge between 2+1+1$d$ and the target model rests on the 
identification of physical fermion fields in 2+1$d$ localised entirely on the domain
walls, which we regard as a
working assumption:
\begin{equation}
\psi(x)=P_-\Psi(x,1)+P_+\Psi(x,L_s);\;\;\;\bar\psi(x)=\bar\Psi(x,L_s)P_-+\bar\Psi(x,1)P_+,
\label{eq:physical}
\end{equation}
with projectors $P_\pm\equiv{1\over2}(1\pm\gamma_3)$.
The mass term $mS_m$ in (\ref{eq:action}) needs some  discussion. A U(2)-symmetric theory has three
physically equivalent parity-invariant mass terms $m_h\bar\psi\psi$,
$im_3\bar\psi\gamma_3\psi$ and $im_5\bar\psi\gamma_5\psi$. For DWF fermions with
$L_s$ finite the choice $im\bar\psi\gamma_3\psi$ with 
$\psi,\bar\psi$ specified in (\ref{eq:physical}) gives the best approach to
$L_s\to\infty$~\cite{Hands:2015qha,Hands:2015dyp,Hands:2018vrd}, and we use this
definition throughout this paper. Approach to the U(2) symmetric limit will be
monitored in Sec.~\ref{sec:Ls48} below via the residual 
$\delta_h\simeq\langle\bar\psi\psi\rangle-\langle\bar\psi
i\gamma_3\psi\rangle$~\cite{Hands:2015qha}.

Specifying the bilinear component of the action (\ref{eq:action}) by ${\cal M}$, 
then the action simulated using bosonic pseudofermion fields
$\Phi,\Phi^\dagger$ is
\begin{equation}
S_{\rm pf}=\Phi^\dagger\left\{[{\cal M^\dagger M}_{m_h=1}]^{1\over4}[{\cal
M^\dagger M}_{m_3=m}]^{-{1\over2}}[{\cal M^\dagger
M}_{m_h=1}]^{1\over4}\right\}\Phi.
\label{eq:Spf}
\end{equation}
Assuming $\mbox{det}{\cal M}$ is real, then the resulting functional weight 
$\mbox{det}[{\cal M}_{m_3=m}{\cal M}^{-1}_{m_h=1}]$ tends to
$\mbox{det}D_{OL}(m)$ in the limit $L_s\to\infty$~\cite{Hands:2015dyp}.
An RHMC algorithm is needed to handle the fractional powers in (\ref{eq:Spf}),
as described in \cite{Hands:2018vrd}. Some tests of the accuracy of
the rational approximation needed to implement fractional powers in the parameter regime of
interest are presented in
Sec.~\ref{sec:Ls48}. 
In the meantime the code has been modified in two main aspects: multiprocess
parallelisation and a simplified, more stable solver.  The parallelisation makes
use of the MPI paradigm, with a 3D domain decomposition across the spatial and
temporal directions, leaving the Domain Wall dimension uncut to allow the
compiler
-- as in the original version of the code -- to perform automatic vectorisation.
  The alternative solver implementation is a conjugate gradient-based multi-shift solver, having
the advantages of requiring 33\% less memory compared to the original QMR-based
one, and being stable in single precision.  The possibility of running the
solver in single precision during the molecular dynamics part of the algorithm
leads to another 50\% save in memory which, as the solver is severely
memory-bound, translates into a direct performance increase.
We have made the simulation code publicly available~\cite{code}.

Finally, the observables studied in the bulk of the paper are simply the bilinear condensate
\begin{equation}
\langle\bar\psi\psi\rangle\equiv{{\partial\ln{\cal Z}}\over{\partial
m}}\equiv\langle\Sigma\rangle
\label{eq:condensate}
\end{equation} 
and its associated susceptibility
\begin{equation}
\chi_\ell=\langle\Sigma^2\rangle - \langle\Sigma\rangle^2.
\label{eq:chi}
\end{equation}
The notation is slightly loose; the bilinear actually used in computations should be understood to be
$i\bar\psi\gamma_3\psi$. The susceptibility $\chi_\ell$ defined in (\ref{eq:chi}) is
often referred to as the {\em disconnected} component, and is expected to
manifest any singular behaviour expected at a continuous phase transition. The
full susceptibility corresponding to $\partial^2\ln{\cal Z}/\partial m^2$
contains an extra {\em connected} component not included in the variance of the
bilinear order parameter calculated here.
In our work the expectations (\ref{eq:condensate},\ref{eq:chi}) are calculated
with unbiassed estimators using 10 independent Gaussian noise vectors per
configuration.

\section{Results}
\label{sec:results}

\subsection{Equation of state on $12^3$, $16^3$, and the Extrapolation
$L_s\to\infty$}
\label{sec:Lsextrap}

In previous work~\cite{Hands:2018vrd}, simulations of the $N=1$ model on a
$12^3\times L_s$ system found signals consistent with broken  U(2) symmetry 
in the vicinity of $\beta\equiv g^{-2}a\approx0.3$. Crucially, in
order to demonstrate this it
proved necessary to explore a large range $8\leq L_s\leq40$ in order to permit
an extrapolation $L_s\to\infty$.  
In this subsection we revisit this issue, this time exploring the
coupling range $0.3\leq\beta\leq0.52$ with much finer resolution, with fermion 
masses $ma=0.01,\ldots,0.05$, spacetime volumes $12^3$ and
$16^3$ and $L_s=8,16,\dots,48$. The data presented result from at least 3000 RHMC
trajectories of mean length 1.0 for each parameter set $\{\beta,m,L_s\}$.

\begin{figure}[tbp]
\begin{center}
  \includegraphics[width=0.75\columnwidth]{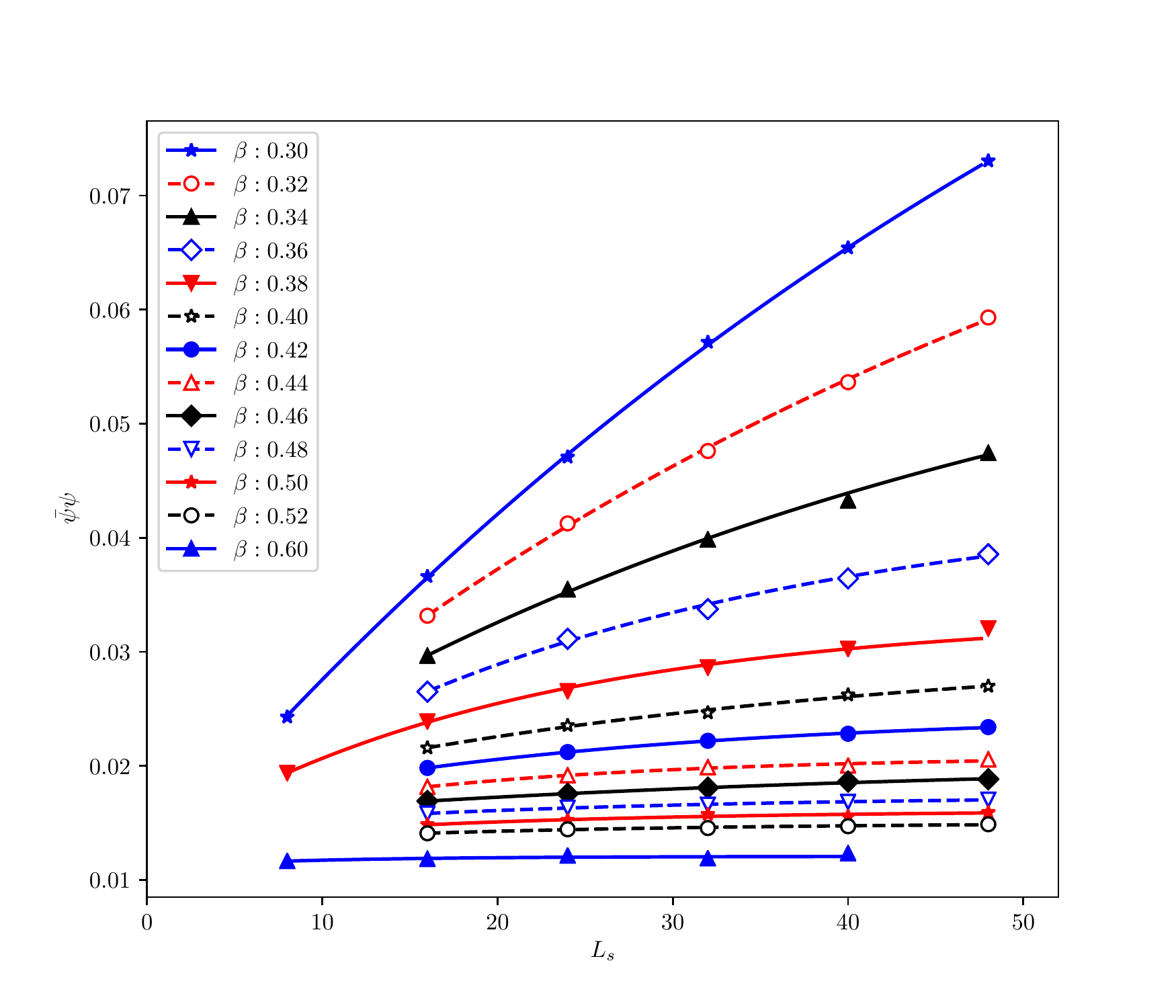}\\
  \caption{Bilinear condensate $\langle\bar\psi\psi(L_s)\rangle$ for $ma=0.01$
on $16^3$ together with a fit to (\ref{eq:Lsextrap}).}
  \label{fig:Lsextrap}
  \includegraphics[width=0.75\columnwidth]{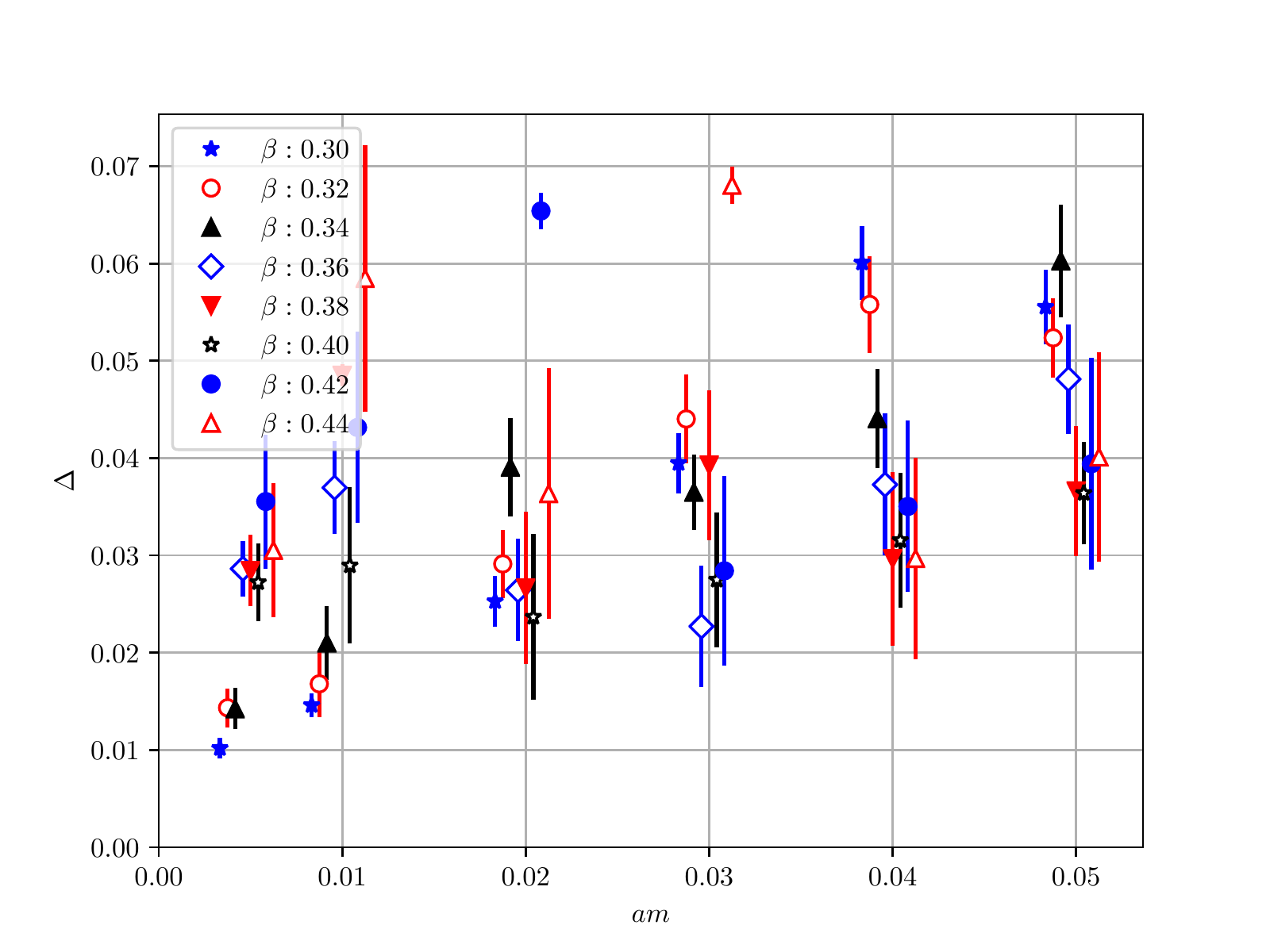}\\
  \caption{The decay constant $\Delta(\beta,m)$ defined in (\ref{eq:Lsextrap}) on
$16^3$, showing distinct trends in the coupling ranges $\beta\geq0.36$ and
$\beta\leq0.34$.
 Data points have been shifted horizontally for improved readability.}
  \label{fig:decay_constants_16}
\end{center}
\end{figure}

First let's discuss the $L_s\to\infty$ extrapolation; empirically the
convergence to the large-$L_s$ limit is exponential~\cite{Hands:2018vrd}:
\begin{equation}
\langle\bar\psi\psi\rangle_{L_s=\infty}-\langle\bar\psi\psi\rangle_{L_s}=
A(\beta,m)e^{-\Delta(\beta,m)L_s}.
\label{eq:Lsextrap}
\end{equation}
Sample fits are shown in Fig.~\ref{fig:Lsextrap}.
The extracted  decay constant $\Delta(\beta,m)$ is shown in
Fig.~\ref{fig:decay_constants_16}.
While errors are appreciable, particularly at weaker couplings where the signal
is small, it can be seen that for the weaker couplings
$0.36\leq\beta\leq0.44$ $\Delta\approx0.03$ and is
approximately $m$-independent,
whereas for the stronger couplings $0.3\leq\beta\leq0.34$, 
$\Delta(m)$ is roughly linear, with an intercept
$\Delta(m=0)\approx0.007$. This suggests that $L_s\sim O(10^2)$ is needed
to completely control the extrapolation in this regime.

Next we turn to analysis of the critical point, assumed present at 
$(\beta,m)=(\beta_c,0)$. Since data are collected in
the presence of a U(2)-symmetry breaking mass $m\not=0$, we follow an 
indirect route, motivated by the renormalisation group~\cite{DelDebbio:1997dv}. 
At fixed spacetime volume, assume a scaling form
\begin{equation}
m(\langle\bar\psi\psi\rangle,t)=\langle\bar\psi\psi\rangle^\delta{\cal
F}(t\langle\bar\psi\psi\rangle^{-{1\over\beta_m}})
\end{equation}
where $t\equiv\beta-\beta_c$. In the limits $m=0$, $t=0$ we immediately recover
\begin{equation}
\vert t\vert^{\beta_m}\propto\langle\bar\psi\psi\rangle;\;\;\;
m={\cal F}(0)\langle\bar\psi\psi\rangle^\delta
\label{eq:crit_exponents}
\end{equation}
whereupon we identify the conventional critical exponents $\beta_m$ and
$\delta$ familiar from the ferromagnetic transition. 
For small $t$ we may Taylor-expand the scaling function ${\cal F}$
to yield the equation of state (EoS):
\begin{equation}
m=A(\beta-\beta_c)\langle\bar\psi\psi\rangle^{\delta-{1\over\beta_m}}+B\langle\bar\psi\psi\rangle^\delta.
\label{eq:eos}
\end{equation}

\begin{figure}[tbp]
\begin{center}
  \includegraphics[width=0.85\columnwidth]{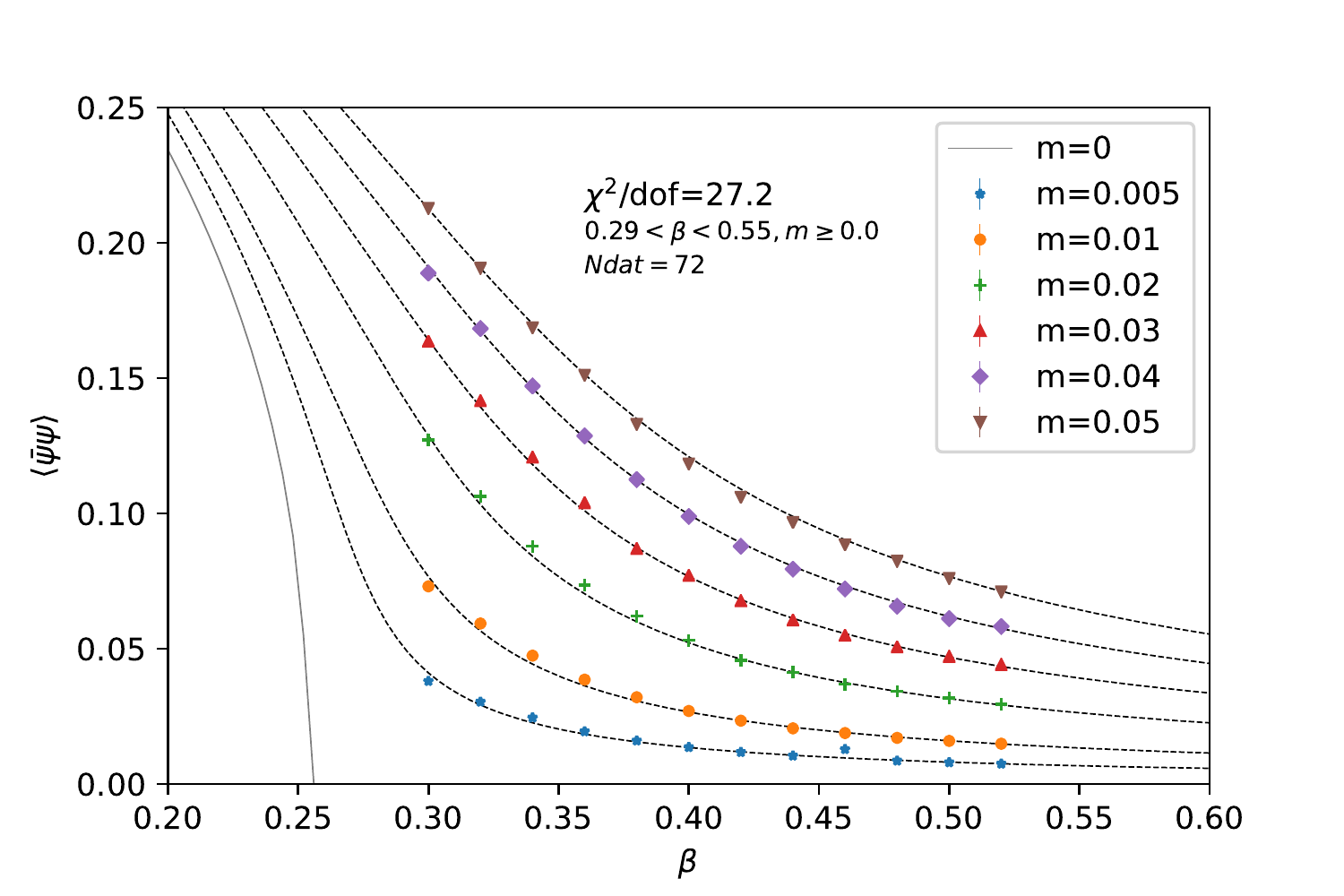}\\
  \caption{Fit to the EoS (\ref{eq:eos}) for $16^3\times48$. The
full line shows the fit extrapolated to the U(2)-symmetric limit $m=0$.
The large $\chi^2$ value is dominated by data with $\beta=0.30$ -- see
Fig.~\ref{fig:eosfit_sc} below.}
  \label{fig:eosfit_16_48}
\end{center}
\end{figure}
Results of a least-squares fit of (\ref{eq:eos}) to data from $16^3\times48$,
$0.3\leq\beta\leq0.52$,
and $ma=0.005$,0.01,0.02,$\ldots,0.05$ are shown in Fig.~\ref{fig:eosfit_16_48},
together with the curve resulting from the same fit parameters as $m\to0$. The
data support a symmetry-breaking continuous phase transition at
$\beta_c=0.2537(2)$ (the  fit of Fig.~\ref{fig:eosfit_sc} below 
with significantly smaller $\chi^2/\mbox{dof}$ results
from excluding the data with $\beta=0.3$; here we retain them for the sake of
uniformity in the subsequent analysis). The fitted exponents $\beta_m=0.42(1)$, $\delta=3.41(5)$ differ 
significantly from their predicted values in mean field theory, namely
$\beta_m={1\over2}$, $\delta=3$.

\begin{figure}[tbp]
\begin{center}
  \includegraphics[width=0.85\columnwidth]{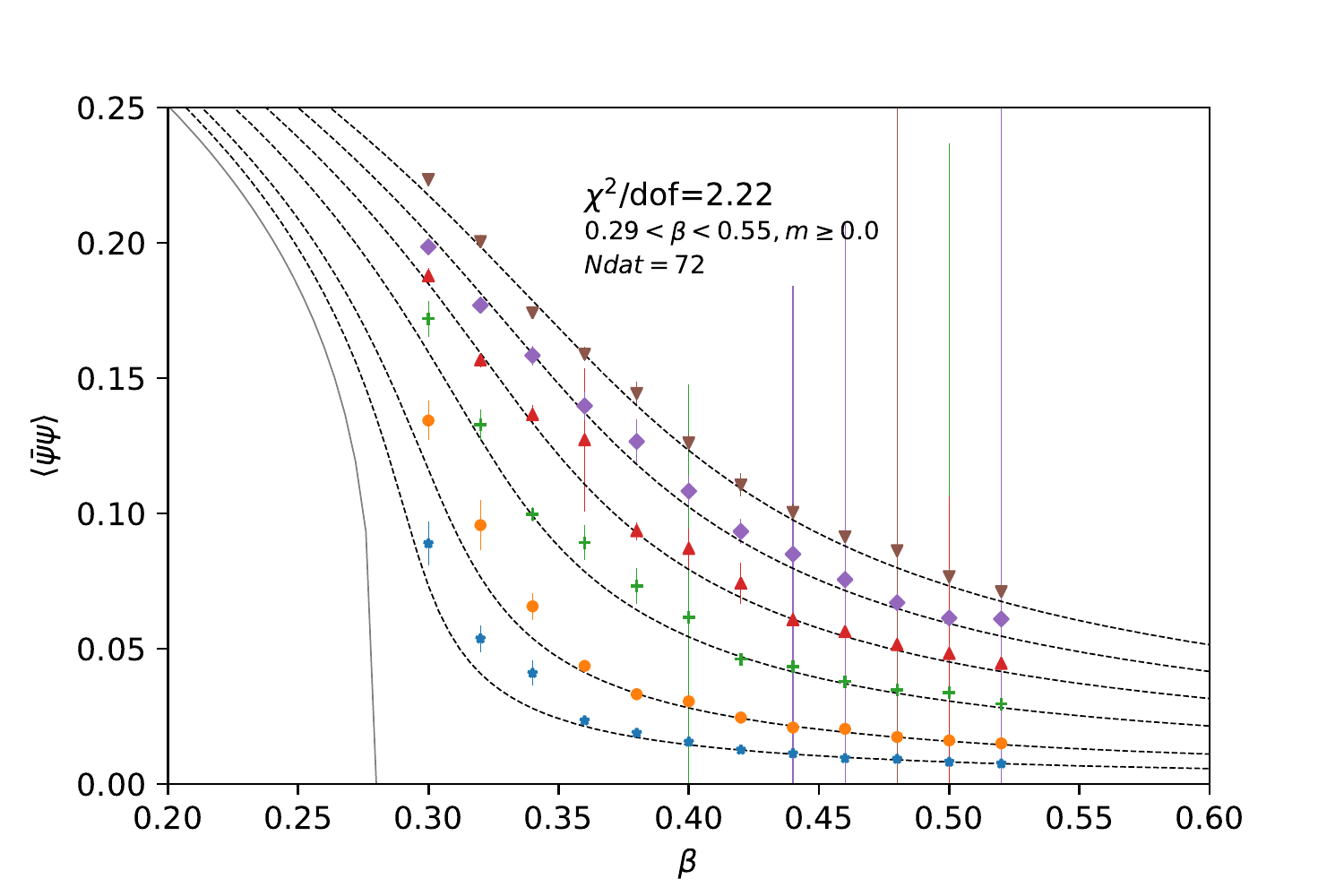}\\
  \caption{Fit to the EoS (\ref{eq:eos}) for $16^3$
extrapolated to infinite $L_s$.}
  \label{fig:eosfit_16_INF}
\end{center}
\end{figure}
However, in order to identify this transition with the breaking of
U(2) symmetry a minimum requirement is stability under the extrapolation
$L_s\to\infty$. The purist's approach, presented first, is to fit (\ref{eq:eos})
to data which has been first extrapolated using (\ref{eq:Lsextrap}). Such a fit
is shown in Fig.~\ref{fig:eosfit_16_INF}. The first thing to note is the far
larger errors on the datapoints (and correspondingly smaller $\chi^2/\mbox{dof}$) -- 
the extrapolation is particularly difficult to control at
weak coupling where the signal is small. The fitted EoS still
supports a continuous phase transition, though with modified exponents
$\beta_m=0.31(2)$, $\delta=4.3(2)$. The critical point is also shifted to weaker 
coupling: $\beta_c=0.279(1)$. In the immediate vicinity of the critical point 
the fitted curves do not well describe the
small-mass data 
$ma=0.005,0.01$; however, due to the large errorbars 
exclusion of these masses does not significantly alter the fit.

\begin{figure}[tbp]
\begin{center}
  \includegraphics[width=0.85\columnwidth,scale=0.8]{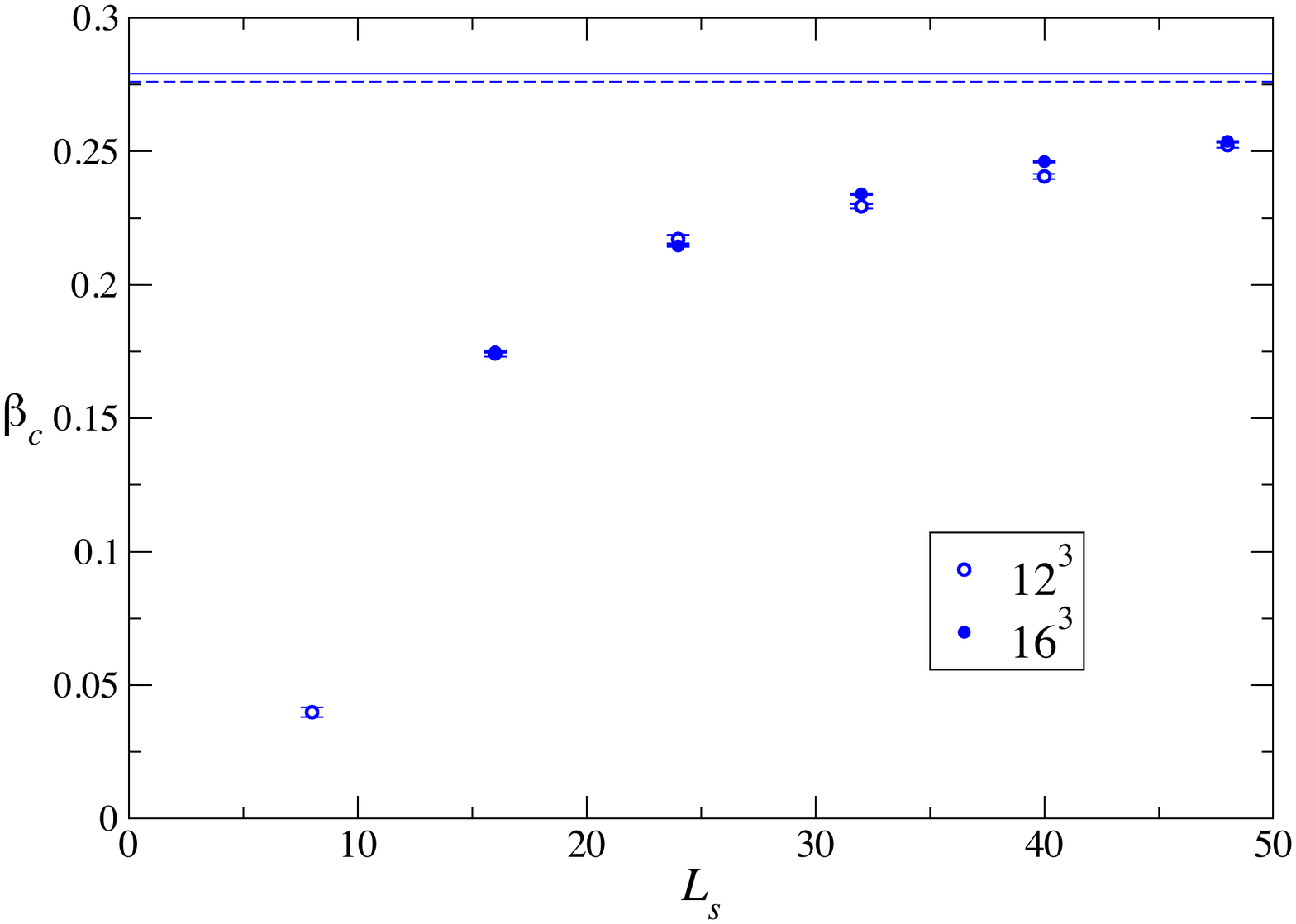}\\
  \caption{Fitted critical coupling $\beta_c$ on both $12^3$ and $16^3$ spacetime
volumes as a function of $L_s$. The full ($16^3$) and dashed ($12^3$) lines show
the values from fits of $L_s\to\infty$ extrapolated data.}
  \label{fig:betac_Ls}
\vspace{1cm}
  \includegraphics[width=0.85\columnwidth,scale=0.8]{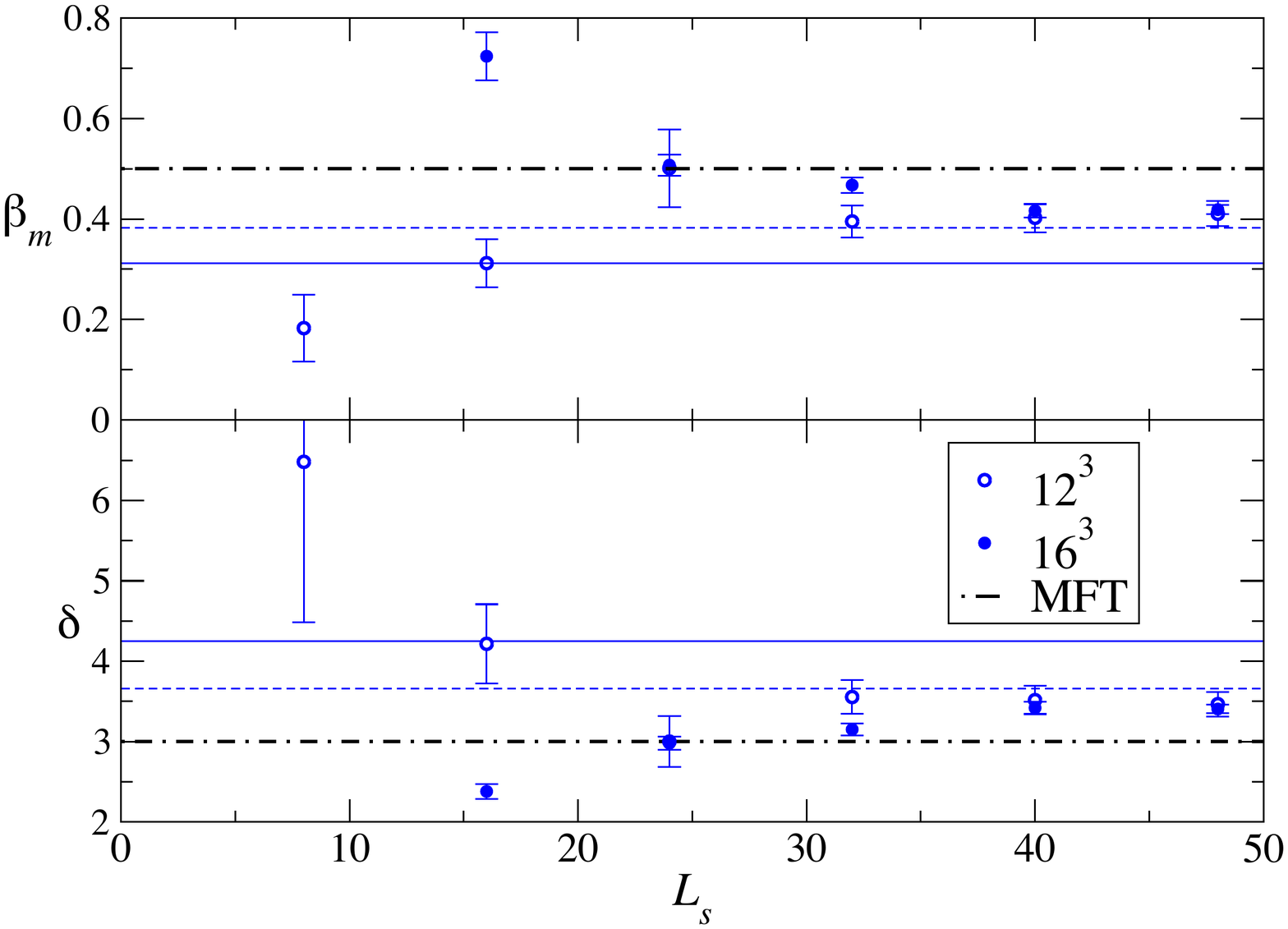}\\
  \caption{Fitted critical exponents $\beta_m$ (upper panel) and $\delta$
(lower panel) on both $12^3$ and $16^3$ spacetime
volumes as a function of $L_s$. Dash-dotted lines show mean field values} 
  \label{fig:exponents_Ls}
\end{center}
\end{figure}
A more pragmatic approach is to perform fits at fixed $L_s$ and then study the behaviour
of the fit parameters as $L_s\to\infty$. This is less well-motivated
theoretically, but has the practical advantage that the data passed to the
least-squares fitting procedure is of much higher statistical quality. Results
for the critical coupling and the exponents are shown in Figs.~\ref{fig:betac_Ls},
\ref{fig:exponents_Ls} respectively. In Fig.~\ref{fig:betac_Ls} it is notable that
$\beta_c(L_s)$ from
$12^3$ and $16^3$ volumes are compatible;  the evolution with $L_s$ is smooth
but falls significantly short of the fit of the extrapolated data even by
$L_s=48$, so it is not clear whether extrapolation and fitting commute.
In Fig.~\ref{fig:exponents_Ls} by contrast the fitted $\beta_m(L_s)$ and
$\delta(L_s)$, while agreeing at large $L_s$, approach this limit from opposite
directions  on $12^3$ and $16^3$; moreover while on
$12^3$ the data from the largest available $L_s$ are compatible with the values
extracted from the extrapolated data, as already remarked there is a significant disparity on $16^3$.

\subsection{Towards Stronger Couplings on $16^3$, $L_s=48$}
\label{sec:Ls48}
The results of the previous subsection support the claim
made in \cite{Hands:2018vrd} that the $N=1$ model has a continuous
phase transition associated with the onset of a bilinear fermion condensate
for $\beta\approx0.3$. There is no sign of any significant finite volume
effect.
Moreover, for the largest $L_s$ examined, 
the transition is reasonably well-modelled by an
EoS with critical exponents both distinct from
mean-field values, and consistent with a QCP defining a previously unknown
strongly coupled quantum field theory of fermions. We might be concerned,
however, that all data analysed in Sec.~\ref{sec:Lsextrap} lie on the
weak-coupling side of the putative transition, and that there are hints from
Fig.~\ref{fig:eosfit_16_48} and particularly Fig.~\ref{fig:eosfit_16_INF} that
the EoS fit is not doing such a good job as either $m\to0$ or
$\beta\to\beta_{c+}$. To investigate further we now turn to data generated using
approximately 5000 RHMC trajectories at
fixed $L_s=48$ on spacetime volume $16^3$, but extending to $\beta$-values on
the strong-coupling side of the transition. Additionally, at the lightest mass $ma=0.005$ the
$\beta$-axis is sampled with a finer resolution $\Delta\beta=0.01$ in the strong
coupling region.

\begin{figure}[tbp]
\begin{center}
  \includegraphics[width=0.85\columnwidth,scale=0.8]{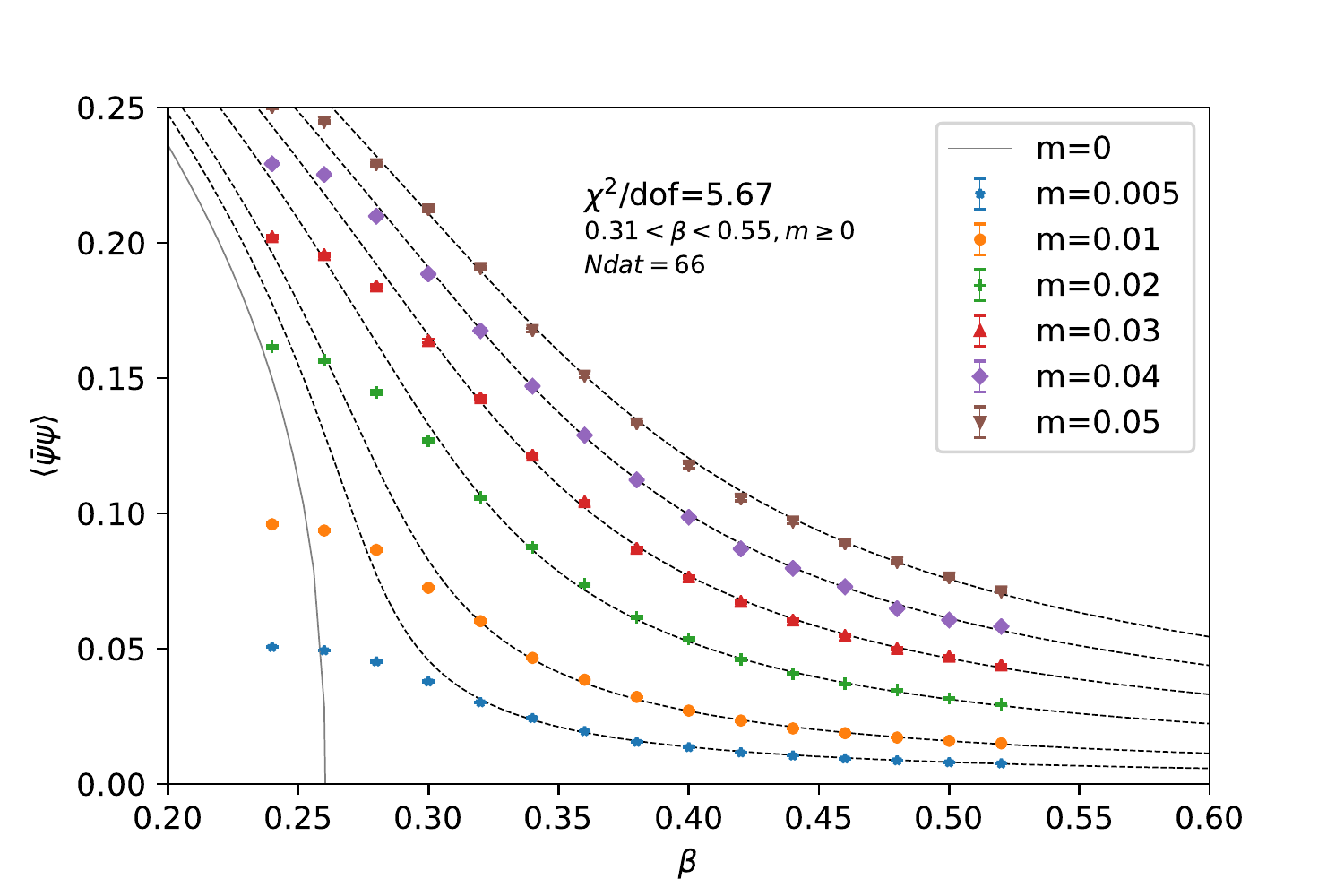}\\
  \caption{The fixed-$L_s$ dataset used for the analysis of
Sec.~\ref{sec:Ls48}, together with a fit to the EoS  
(\ref{eq:eos}). The fit parameters are $\beta_c=0.2601(4)$, $\beta_m=0.413(15)$,
$\delta=3.44(9)$.}
  \label{fig:eosfit_sc}
\end{center}
\end{figure}
\begin{figure}[tbp]
\begin{center}
  \includegraphics[width=0.85\columnwidth,scale=0.8]{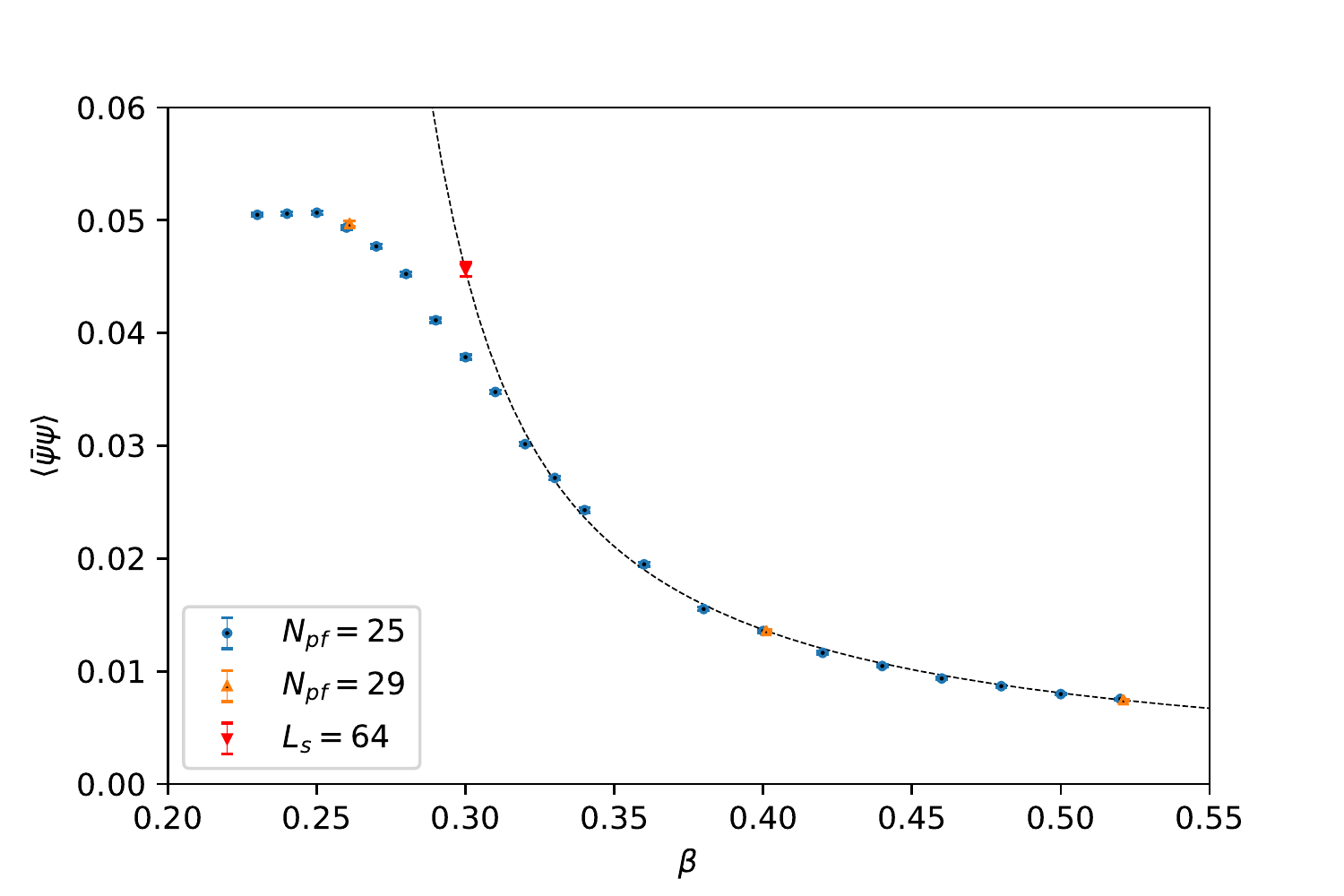}\\
  \caption{$\langle\bar\psi\psi(\beta)\rangle$ for $ma=0.005$ on $16^3\times48$.
The dashed line
is the same EoS fit shown in Fig.~\ref{fig:eosfit_sc}. 
Also shown is the result 
of a pilot simulation with $\beta=0.3$, $L_s=64$.}
  \label{fig:cond_light}
\end{center}
\end{figure}
Fig.~\ref{fig:eosfit_sc} shows the $\langle\bar\psi\psi(\beta,m)\rangle$
dataset, and a fit to (\ref{eq:eos}) based on $0.32\leq\beta\leq0.52$. The fit is
compatible with that shown in Fig.~\ref{fig:eosfit_16_48}, and as advertised is
of slightly higher quality as a result of excluding $\beta=0.30$. It is
immediately apparent, however, that it fails to model data on the strong
coupling side of the transition; here $\langle\bar\psi\psi\rangle$ falls below
the model, the effect is more pronounced with decreasing $m$, until by
$ma=0.005$ the curve becomes flat for $\beta\lapprox0.25$, as shown in
Fig.~\ref{fig:cond_light}.

The flattening of the condensate at strong coupling is not a new story; indeed, simulations with
staggered fermions actually exhibit a maximum before dropping as
$\beta\to0_+$. A possible origin of this behaviour was
suggested in \cite{DelDebbio:1997dv}: as a result of the linear coupling between $A_\mu$ and the fermion
current, the leading order large-$N$ correction to the auxiliary propagator
is not transverse in a lattice regularisation, leading to breakdown of
positivity as $\beta\to0$. In the large-$N$ approach the effect is mitigated by
an additive renormalisation of $g^{-2}$, so that the strong coupling limit of the
model is now taken as $\beta\to\beta^*>0$. In \cite{Christofi:2007ye} 
$\beta^*$ was taken to be the location of the maximum of
$\langle\bar\psi\psi(\beta)\rangle$, enabling a model equation of state 
$\langle\bar\psi\psi(N)\rangle$ in the effective strong-coupling limit and consequent
prediction of $N_c$. A decrease of $\langle\bar\psi\psi\rangle_{\beta\to0}$ 
has also been reported in simulations with SLAC fermions
\cite{Wellegehausen:2017goy,Lenz:2019qwu} and with DWF in a variant ``surface''
formulation of the model~\cite{Hands:2016foa}, suggesting that strong coupling lattice
artifacts are a generic feature of the Thirring model, and may have a more
general origin than that suggested by the large-$N$ approach.
Be that as it may it will clearly be important to establish a clear
separation between $\beta^*$ and any $\beta_c$ associated with a Thirring model
QCP. From Fig.~\ref{fig:cond_light} we might estimate
$\beta^*\approx0.25$, uncomfortably close, with current resolution, to the
$\beta_c$ estimates of Sec.~\ref{sec:Lsextrap}.

At this point it is appropriate to discuss a technical aside. In the RHMC
algorithm described in \cite{Hands:2018vrd}, it is necessary to calculate
fractional powers of the fermion kernel ${\cal A=M^\dagger M}$. In practice this
is performed using a rational approximation
\begin{equation}
{\cal A}^p\simeq r_p({\cal A})=\alpha_0+\sum_{i=1}^{N_{pf}}{\alpha_i\over{{\cal
A}+\beta_i}},
\end{equation}
where the coefficients $\alpha_i,\beta_i$ may be calculated using the Remez
algorithm implementation described in \cite{ClarkKennedy}. They are chosen so
that over a spectral range $(\lambda_d,50.0)$, $\vert r_p(x)-x^p\vert<10^{-6}$
for matrices needed during trajectory guidance and $<10^{-13}$ for those needed
in the Monte Carlo acceptance step. For all work to date we have used $\lambda_d=10^{-4}$
corresponding to the smallest value of $(ma)^2$ explored, which translates to 
partial fraction numbers $N_{pf}=12$ (guidance) and $N_{pf}=25$ (acceptance); however one might
question whether this is sufficiently accurate for studies with $ma=0.005$.
Accordingly we have performed ``enhanced'' simulations at three $\beta$-values
with Remez coefficients generated with $\lambda_d=10^{-5}$, corresponding to 
$N_{pf}=14$ (guidance), and $N_{pf}=29$ (acceptance). As shown in
Fig.~\ref{fig:cond_light}, fortunately there appears to be no significant
difference with data calculated using the previous $\lambda_d=10^{-4}$.

\begin{figure}[tbp]
\begin{center}
  \includegraphics[width=0.85\columnwidth,scale=0.8]{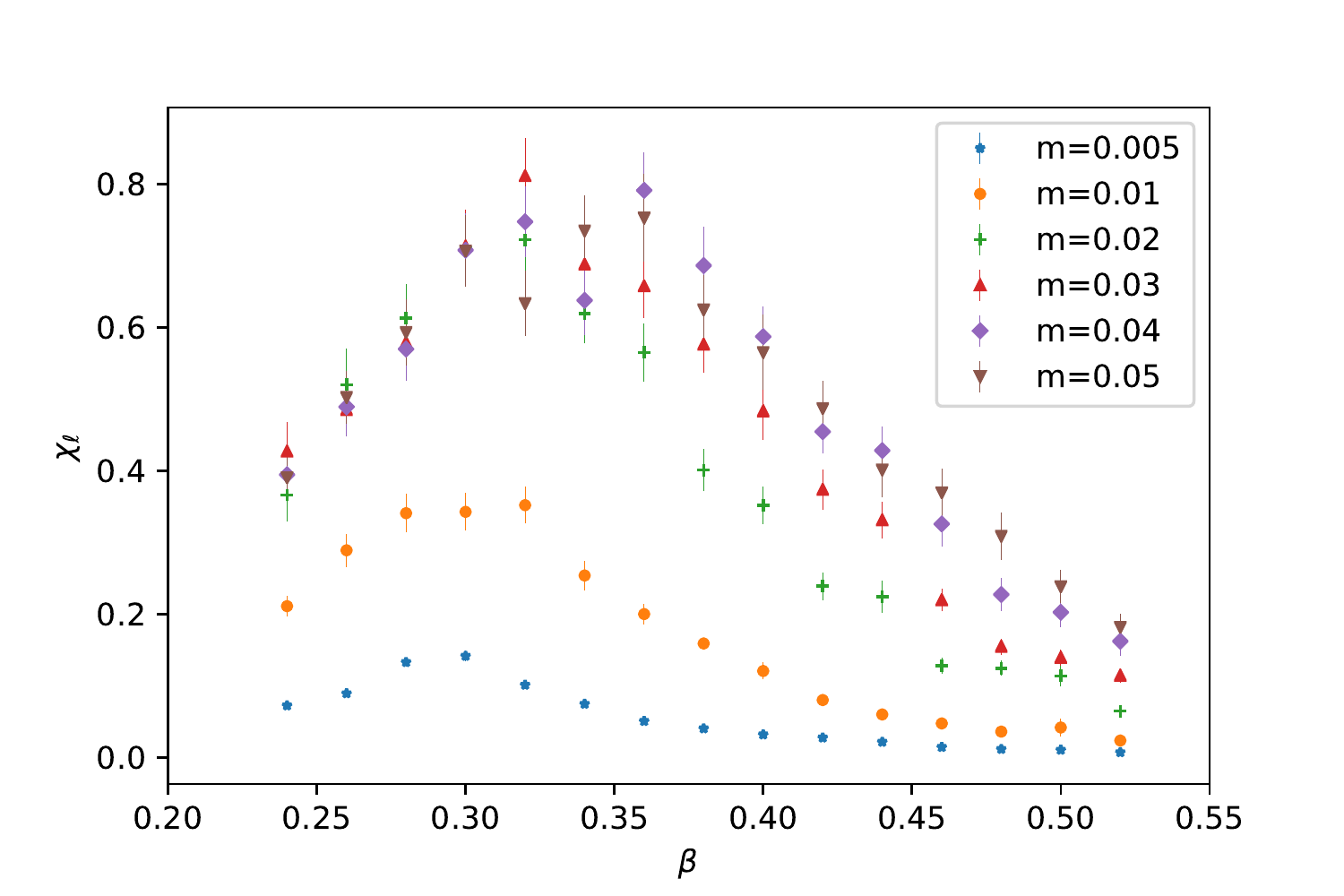}\\
  \caption{Susceptibility $\chi_\ell(\beta,m)$ on $16^3\times48$.}
  \label{fig:suscpeek}
  \includegraphics[width=0.85\columnwidth,scale=0.8]{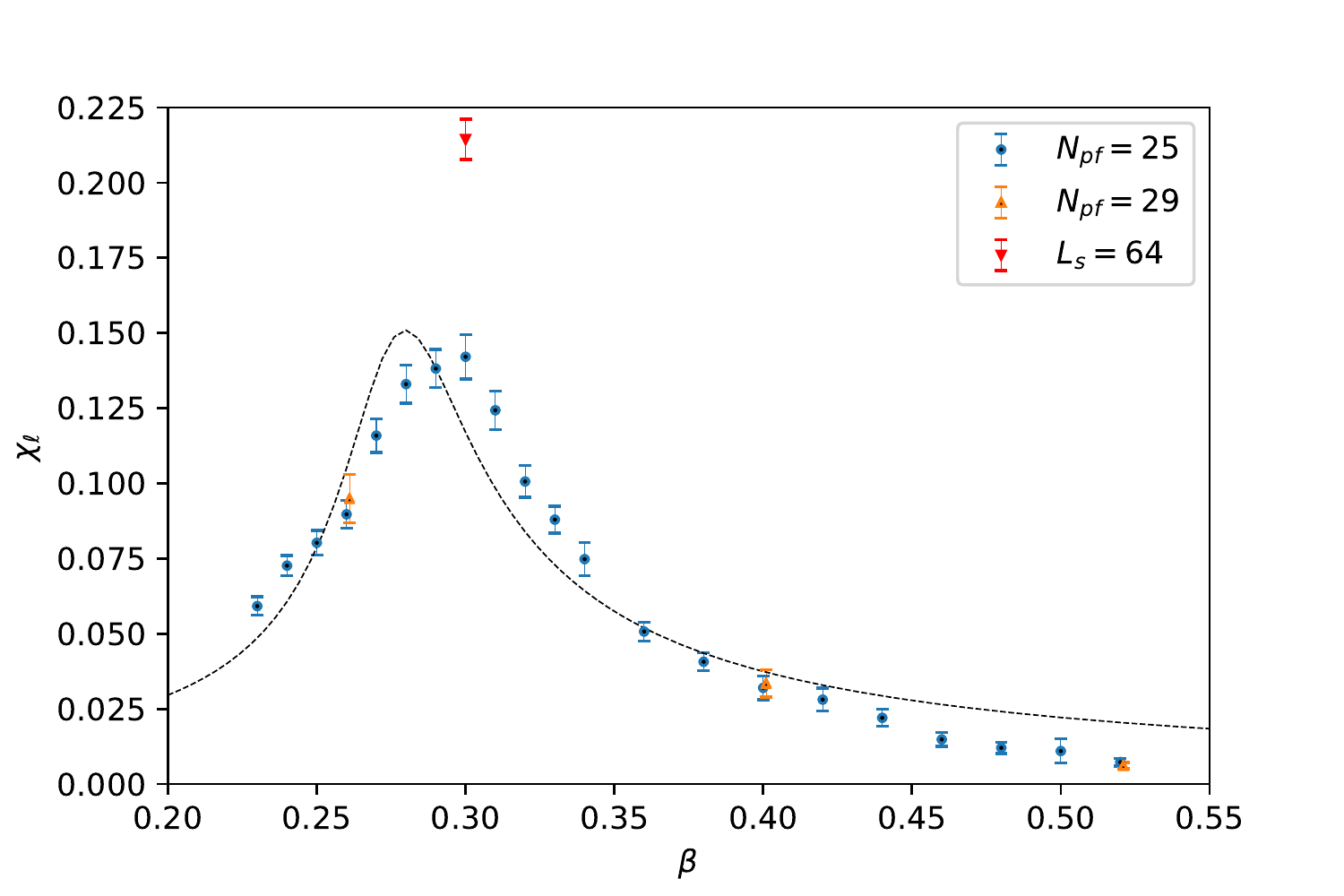}\\
  \caption{Susceptibility $\chi_\ell(\beta)$ for $ma=0.005$ on $16^3\times48$. 
The dashed line is calculated using
the same EoS fit shown in Fig.~\ref{fig:eosfit_sc},
multiplied by an empirical factor 0.014.
Also shown is the result 
of a pilot simulation with $\beta=0.3$, $L_s=64$.}
  \label{fig:susc_light}
\end{center}
\end{figure}
\begin{figure}[tbp]
\begin{center}
  \includegraphics[width=0.85\columnwidth,scale=0.8]{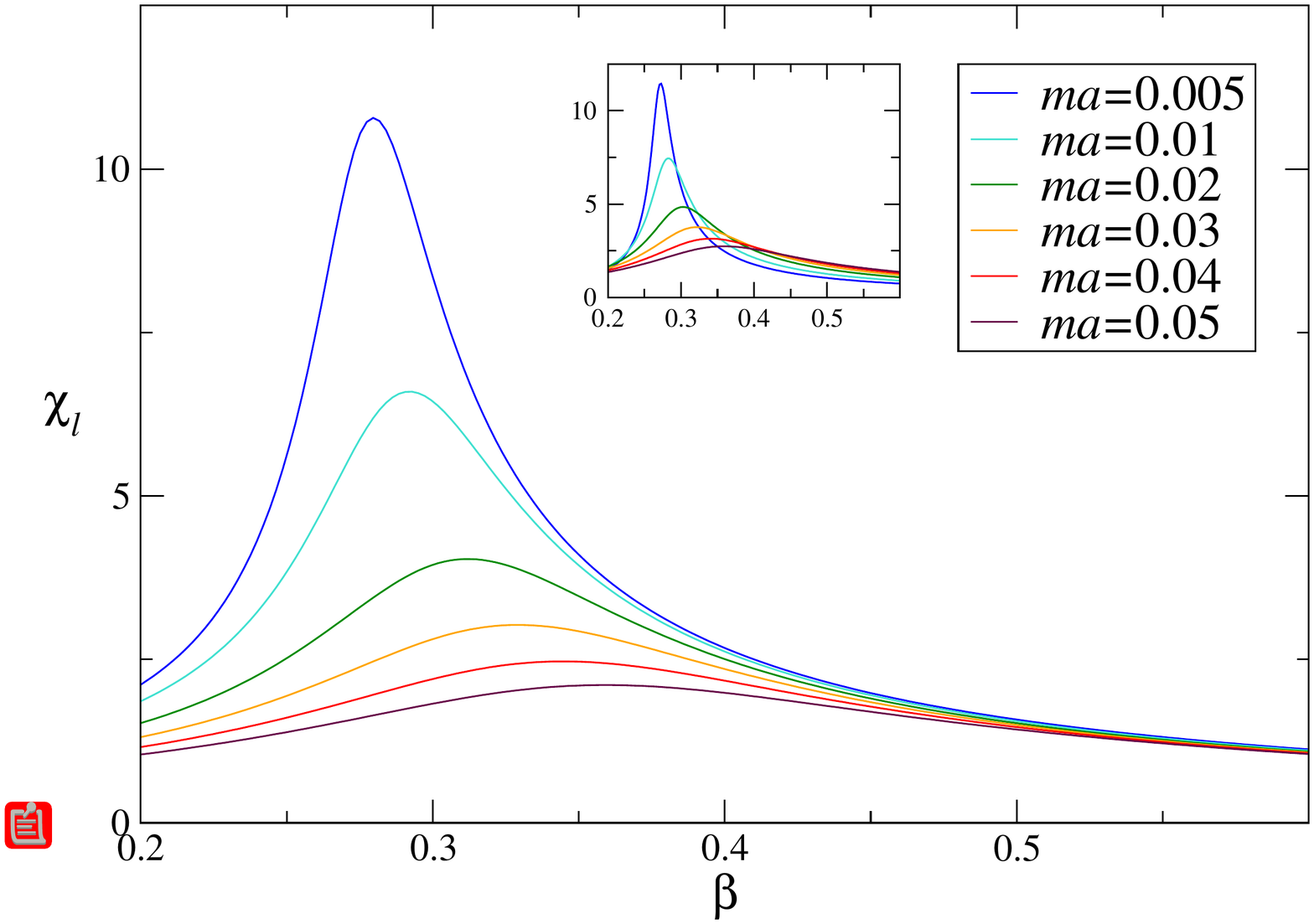}\\
  \caption{Susceptibility $\chi_\ell(\beta,m)$ obtained by differentiation of
the EoS (\ref{eq:eos}). The inset shows the ``corrected'' version
discussed in Sec.~\ref{sec:discussion}}
  \label{fig:model_susc}
\end{center}
\end{figure}

Next we present data for the susceptibility $\chi_\ell$ defined in
(\ref{eq:chi}), for the whole dataset in Fig.~\ref{fig:suscpeek} and for the
lightest $ma=0.005$ in Fig.~\ref{fig:susc_light}.
As might be anticipated, statistical errors in $\chi_\ell$ are considerably
larger than those for the condensate, and accordingly we choose not to attempt
an $L_s\to\infty$ extrapolation. However, 
again, the agreement between results obtained using the default and 
enhanced rational approximations seen in Fig.~\ref{fig:susc_light} is reassuring.
For each value of $m$ $\chi_\ell(\beta)$ is non-monotonic, the peak 
shifting to stronger coupling as $m$ decreases in accord 
with expectations for a second derivative of the free energy at a critical
point; this is corroborated by the model prediction obtained by differentiation
of (\ref{eq:eos}) with respect to $m$, and plotted using the fitted parameters
in Fig.~\ref{fig:model_susc}.
Fig.~\ref{fig:susc_light} suggests that the location and even the sharpness of the peak at
criticality is roughly as expected, once an empirical rescaling is applied. 
However, there are features of Fig.~\ref{fig:suscpeek} which are clearly problematic;
the $m$-ordering of the data is opposite to model expectations, with $\chi_\ell$
{\em increasing\/} with $m$ over the whole $\beta$-range studied, and the
convergence of $\chi_\ell$  curves with different $m$ as $\beta$ grows large
seen in Fig.~\ref{fig:model_susc} is {\em not\/} observed.
We postpone further discussion of these issues to Sec.~\ref{sec:discussion}.

\begin{figure}[tbp]
\begin{center}
  \includegraphics[width=0.8\columnwidth,scale=0.8]{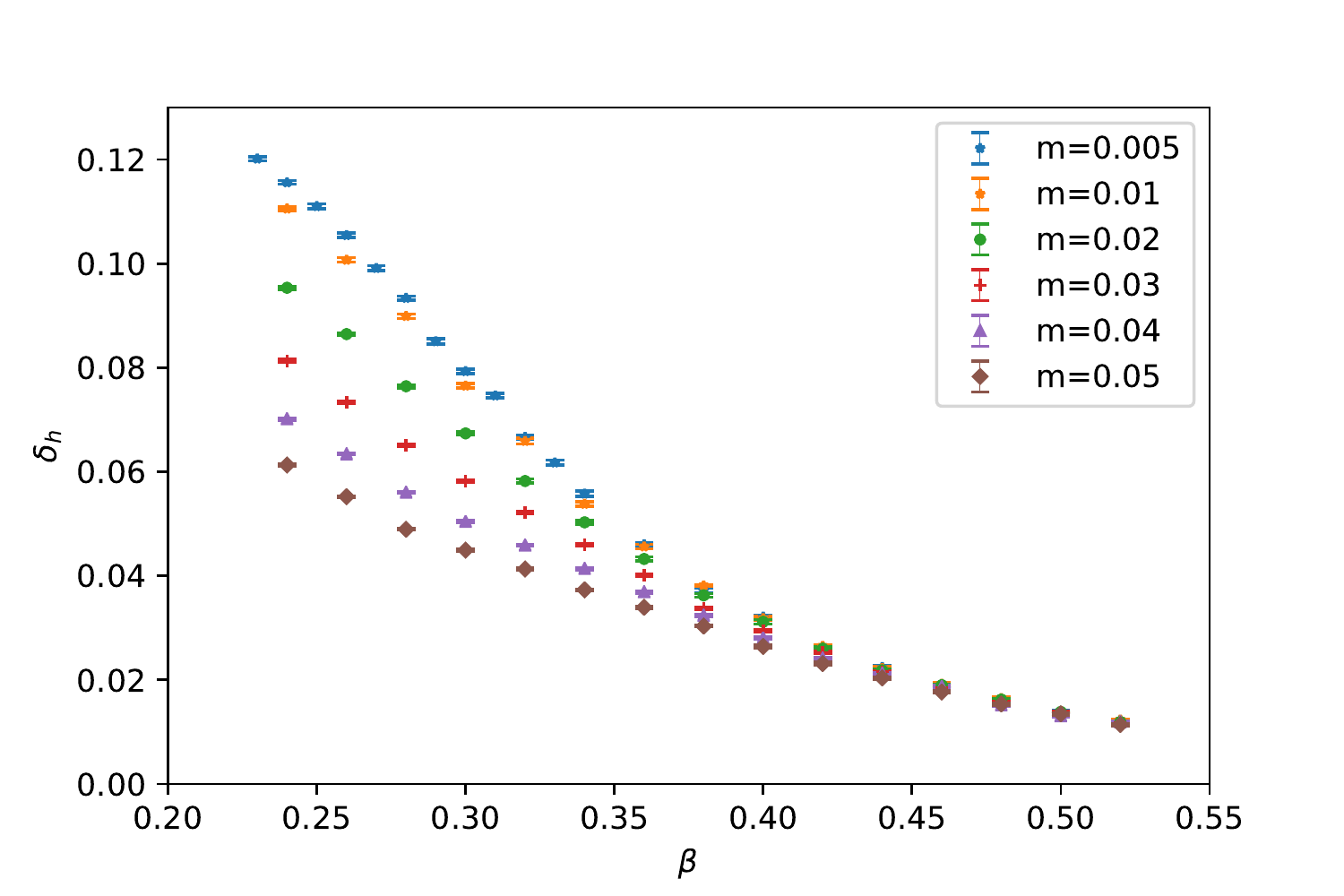}\\
  \caption{The U(2)-breaking residual $\delta_h(\beta,m)$ on $16^3\times48$.}
  \label{fig:deltapeek}
\end{center}
\end{figure}
\begin{figure}[tbp]
\begin{center}
  \includegraphics[width=0.85\columnwidth,scale=0.8]{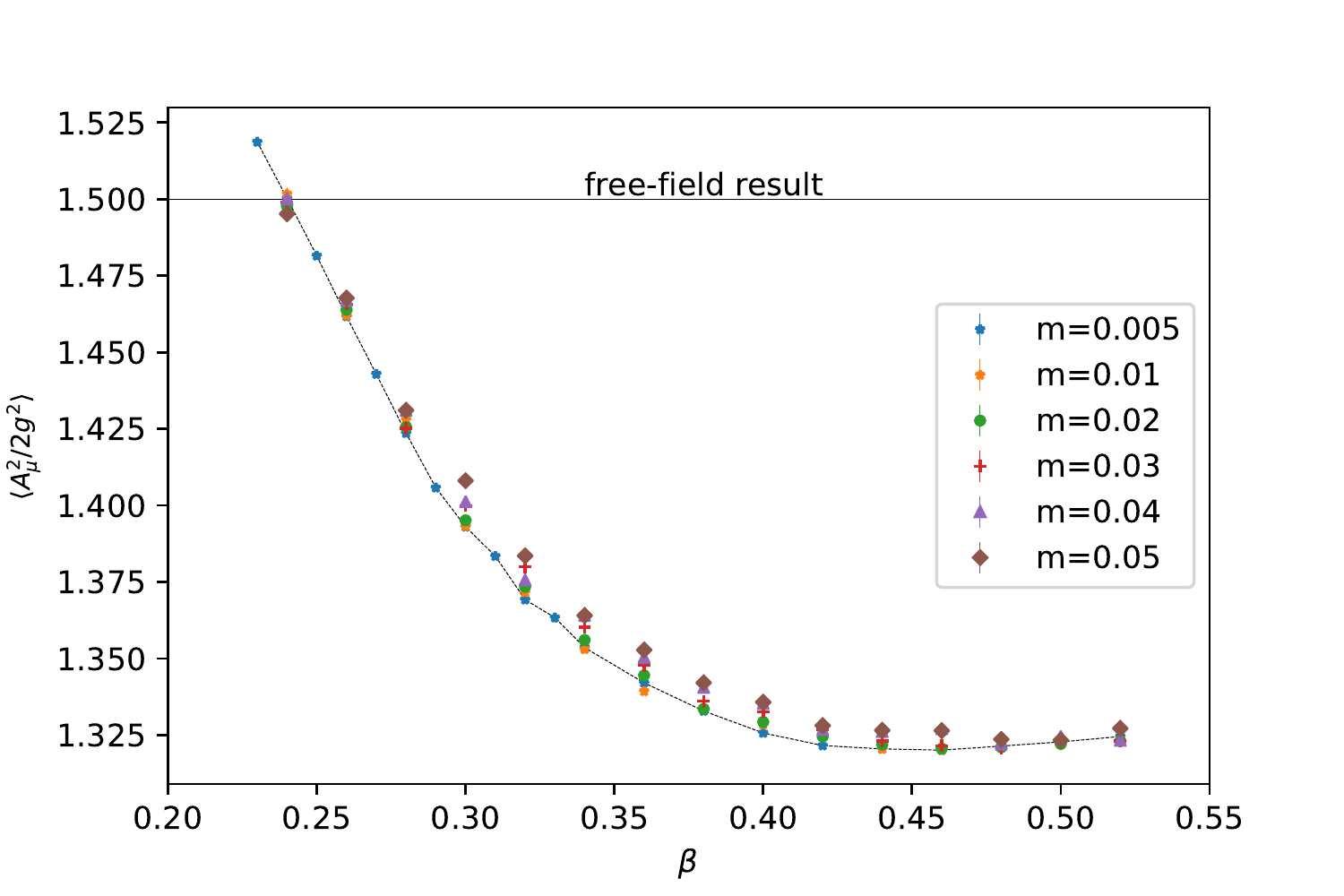}\\
  \caption{Auxiliary action density $(2g^2)^{-1}\langle A_\mu^2\rangle$ vs. $\beta$
on $16^3\times48$.
The dashed line through the $ma=0.005$ data is merely to guide the eye.}
  \label{fig:bosepeek}
\end{center}
\end{figure}
Next we discuss the approach to recovery of U(2) symmetry expected as
$L_s\to\infty$. Ref.~\cite{Hands:2015qha} introduced a residual $\delta_h$
defined in terms of the 2+1+1 dimensional fields as follows:
\begin{equation}
\delta_h(L_s)=\mbox{Im}\langle\bar\Psi(x,s=1)i\gamma_3\Psi(x,s=L_s)\rangle
=-\mbox{Im}\langle\bar\Psi(x,s=L_s)i\gamma_3\Psi(x,s=1)\rangle.
\label{eq:deltah}
\end{equation}
In \cite{Hands:2015qha} $2\delta_h$ was found to furnish a lower bound for the
difference between $\langle\bar\psi\psi\rangle$ and $\langle\bar\psi
i\gamma_3\psi\rangle$, and to vanish $\propto e^{-cL_s}$ for quenched QED$_3$.
In the Thirring model, $\delta_h(L_s)$ for various couplings was presented in
Fig.~12 of \cite{Hands:2018vrd}. While in all cases $\delta_h$ still decreases
with $L_s$, it grows larger as coupling increases, and by $\beta=0.3$
its decay constant $c$ even develops a dependence on $m$.
Fig.~\ref{fig:deltapeek} taken at fixed $L_s=48$  confirms that 
$m$-dependence of $\delta_h$ does indeed set in for $\beta\lapprox0.4$, 
and that $\delta_h$ continues to grow as $\beta$ decreases, suggesting
that recovery of U(2) symmetry will be an ever-increasing challenge in the
symmetry broken phase as $m\to0$.

Finally, Fig.~\ref{fig:bosepeek} shows results for the bosonic auxiliary action
density $(2g^2)^{-1}\langle A_\mu^2\rangle$. As discussed in
\cite{Hands:2018vrd}, for DWF with the bulk formulation of the
Thirring model there is no simple
interpretation in terms of a local four-fermion condensate available; rather we
regard it as an extra observable sensitive to light fermion
dynamics.  Its behaviour is non-monotonic, with a minimum at $\beta\simeq0.46$
before rising to approach and then exceed the free-field value ${3\over2}$ at
$\beta\simeq0.24$. The notable feature of Fig.~\ref{fig:bosepeek} is the
fermion mass-dependence; broadly speaking the departure from the free-field
result increases with
decreasing $m$ (although the $m$-ordering of the data is somewhat noisy), 
the effect being most pronounced for 
$0.3\lapprox\beta\lapprox0.4$ immediately above the suspected critical
region.

\subsection{Properties  of the Associated Overlap Operator }
\label{sec:locality}

The equivalence of DWF~\cite{Kaplan:1992bt,Shamir:1993zy} and the (truncated) overlap
operator~\cite{Neuberger:1997fp} is well established 
in 3+1$d$, eg. \cite{Brower:2012vk}. This equivalence is
further shown in 2+1$d$~\cite{Hands:2015dyp} for both the regular mass term
$m\bar\psi\psi$ and the
linearly independent twisted mass terms $im\bar\psi\gamma_{3,5}\psi$ introduced above. 
As such, locality of
the domain wall operator in the target dimensionality can be demonstrated by showing the locality of the
overlap operator.

We use the Shamir and Wilson formulations of the overlap operator with twisted
mass $-im\bar\psi\gamma_{3}\psi$ given by 
\begin{equation}\label{eq:dol}
D^{3}_{OL}(m)=\frac{1-im\gamma_3}{2}+\frac{1+im\gamma_3}{2}V_{S/W}
\end{equation} 
where the Wilson and Shamir kernels, defined via $V=\gamma_3\mbox{sgn}(H)$, are
\begin{equation}\label{eq:wilsha}
\begin{split} H_W & =\gamma_3 D_W \\ H_S & =
\gamma_3 \frac{D_W}{2+D_W} \\ \end{split} 
\end{equation} 
and
$D_W\equiv D_W(-M)$ is the massive Wilson Dirac operator, $M$ fixed to 1.  
Note, however, that the link factors multiplying the difference operators in
$D_W$ are the non-unitary $[1\pm iA_\mu]$ rather than the unitary $e^{\pm iA_\mu}$
characteristic of a gauge theory. The key relation $H\gamma_3=(\gamma_3H)^\dagger$
is preserved. Our formulation of DWF in the $L_s\to\infty$ limit is expected to
recover (\ref{eq:dol}) with the Shamir kernel~\cite{Hands:2015dyp}.

The
standard mass formulation of the overlap operator is
$D^I_{OL}(m)=\frac{1+m}{2}+\frac{1-m}{2}V_{S/W}$. The signum function typically
is approximated with a rational function resulting in a {\rm truncated\/} overlap
operator. The hyperbolic tangent (polar) approximation is the most commonly used
and may be expressed as \cite{Kennedy:2006ax} 
\begin{equation}\label{eq:polar} 
\text{sgn}(x)  \approx\text{tanh}(n\text{tanh}^{-1}x)   = xn
\frac{\prod_{j=1}^{n/2-1}[x^2+(\text{tan}
\frac{j\pi}{n})^2]}{\prod_{j=0}^{n/2-1}[x^2+(\text{tan}\frac{(j+1/2)\pi}{n})^2]}
\end{equation} 
for $n$ even. For the Shamir formulation it is much
more efficient to use a formulation exploiting an extra dimension, and the
truncated overlap operator can be reconstructed directly from a domain wall
formulation. The standard mass and alternative mass domain wall operators may be
expressed as $D_{DW}^{I/3}=D_{DW}^0+mD_{DW}^{I/3}$, where (for $n\equiv L_s=4$)
\begin{equation}
D_{DW}^0 = \begin{pmatrix} D_W+I & -P_- & 0 & 0 \\ -P_+ & D_W+I & -P_- & 0 \\ 0
& -P_+ & D_W+I & -P_- \\ 0 & 0 & -P_+ & D_W+I \end{pmatrix} \end{equation}
\begin{equation} D_{DW}^I = \begin{pmatrix} 0 & 0 & 0 & P_+ \\ 0 & 0 & 0 & 0 \\
0 & 0 & 0 & 0 \\ P_- & 0 & 0 & 0 \end{pmatrix}, D_{DW}^{3} = \begin{pmatrix} 0
& 0 & 0 & i \gamma_3 P_+ \\ 0 & 0 & 0 & 0 \\ 0 & 0 & 0 & 0 \\ i\gamma_3 P_- & 0
& 0 & 0 \end{pmatrix} \end{equation} Then with \begin{equation} C =
\begin{pmatrix} P_- & P_+ & 0 & 0 \\ 0 & P_- & P_+ & 0 \\ 0 & 0 & P_- & P_+ \\
P_+ & 0 & 0 & P_- \\ \end{pmatrix}, C^\dagger = C^{-1} = \begin{pmatrix} P_- & 0
& 0 & P_+ \\ P_+ & P_- & 0 & 0 \\ 0 & P_+ & P_- & 0 \\ 0 & 0 & P_+ & P_- \\
\end{pmatrix} 
\end{equation} 
we have the following relation
\cite{Hands:2015dyp,Brower:2012vk}, where the precise form of the $\triangle$ terms is
unimportant for our purposes.  
\begin{equation}\label{eq:k} K^{3}=C^\dagger
(D^I_{DW}(1))^{-1}D^{3}_{DW}(m)C = \begin{pmatrix} D^{3}_{OL}(m) & 0 & 0 & 0
\\ -(1-m)\triangle_2^R & 1 & 0 & 0 \\ -(1-m)\triangle_3^R & 0 & 1 & 0 \\
-(1-m)\triangle_4^R & 0 & 0 & 1 \\ \end{pmatrix} 
\end{equation} 
So the overlap
operator (\ref{eq:dol}), with the Shamir kernel (\ref{eq:wilsha}), truncated
with the polar approximation (\ref{eq:polar}), evaluated with given $n$, is
identical to the top left entry of matrix $K$ (\ref{eq:k}) using domain wall
extent $L_s=n$. There is a similar relation for the Wilson kernel, with a
different domain wall formulation. We also have $K^I=C^\dagger
(D^I_{DW}(1))^{-1}D^I_{DW}(m)C$.

Invariant transformations, corresponding 
to the Ginsparg-Wilson (GW) relations (\ref{eq:GW}) are given
by~\cite{Hands:2015qha}
\begin{equation}
\begin{split}
\Psi\to e^{i\alpha\gamma_3(1-\frac{aD}{2})}\Psi & \; ; 
\; \bar{\Psi}\to \bar{\Psi}e^{i\alpha\gamma_3(1-\frac{aD}{2})} \\
\Psi\to e^{i\alpha\gamma_5(1-\frac{aD}{2})}\Psi & \; ; 
\; \bar{\Psi}\to \bar{\Psi}e^{i\alpha\gamma_5(1-\frac{aD}{2})} \\
\Psi\to e^{i\alpha\gamma_3\gamma_5}\Psi & \; ; \; \bar{\Psi}\to \bar{\Psi}e^{i\alpha\gamma_3\gamma_5} \\
\end{split}
\end{equation}
This relation is exact for the 2+1$d$ overlap operator $D=D_{OL}$, and is reproduced
by DWF in 2+1+1$d$ 
in the $L_s\to\infty$ limit. In order to recover the $U(2)$ symmetry in
the continuum limit $a\to 0$, we must have the GW terms $aD\gamma_{3,5}D$ and
equivalently the transform terms $\frac{aD}{2}$ vanishing in the same limit. A
sufficient condition for this to be the
case is the Dirac operator being exponentially local. Evidence for this has been
given for the overlap operator in 3+1$d$~\cite{Hernandez:1998et}, and it behooves us to investigate
the 2+1$d$ case. Since the overlap operator is a dense matrix and manifestly
non-local, 
demonstration of exponential locality is important,
especially around critical regions. 

To see that exponential locality is necessary to ensure recovery of the
continuum $U(2)$ symmetry, note that
\begin{equation}
e^{i\alpha\gamma_3(1-\frac{aD}{2})}\Psi = (I+i\alpha\gamma_3(1-\frac{aD}{2}) + \cdots) \Psi
\end{equation}
so that recovery requires  
\begin{equation}
[aD\Psi]_{a\to 0}=0
\end{equation}
We have $\Psi^\prime_j=[a\sum_iD_{ji}\Psi_i]_{a\to 0}=0$ which is true if
$[\sum_i D_{ji} \Psi_i]_{a\to 0} < \infty$ which is true for any bounded $\Psi$
if $[\sum_i D_{ji}]_{a\to 0} < \infty$ which is true if $D$ is exponentially
local, and hence exponential locality allows recovery of $U(2)$ symmetry.

In order to illustrate the locality of the overlap operator, we follow
\cite{Hernandez:1998et}. Let 
\begin{equation}
\psi(x) = D\eta(x)
\end{equation}
where the point source $\eta(x)=\delta_{x,y}\delta_{\alpha,1}$ where $y$ is an arbitrary location
and $\alpha$ a spinor index. Then we calculate the decay as
\begin{equation}
f(r)=\text{max}\{||\psi(x)||_2 : ||x-y||_1 = r\}
\end{equation}
where the ``Manhattan taxi distance'', $||x-y||_1 = \sum_\mu |x_\mu -y_\mu|$ is just the $L_1$ norm. 

\begin{figure}[tbp]
\begin{center}
  \includegraphics[width=0.85\columnwidth,scale=0.8]{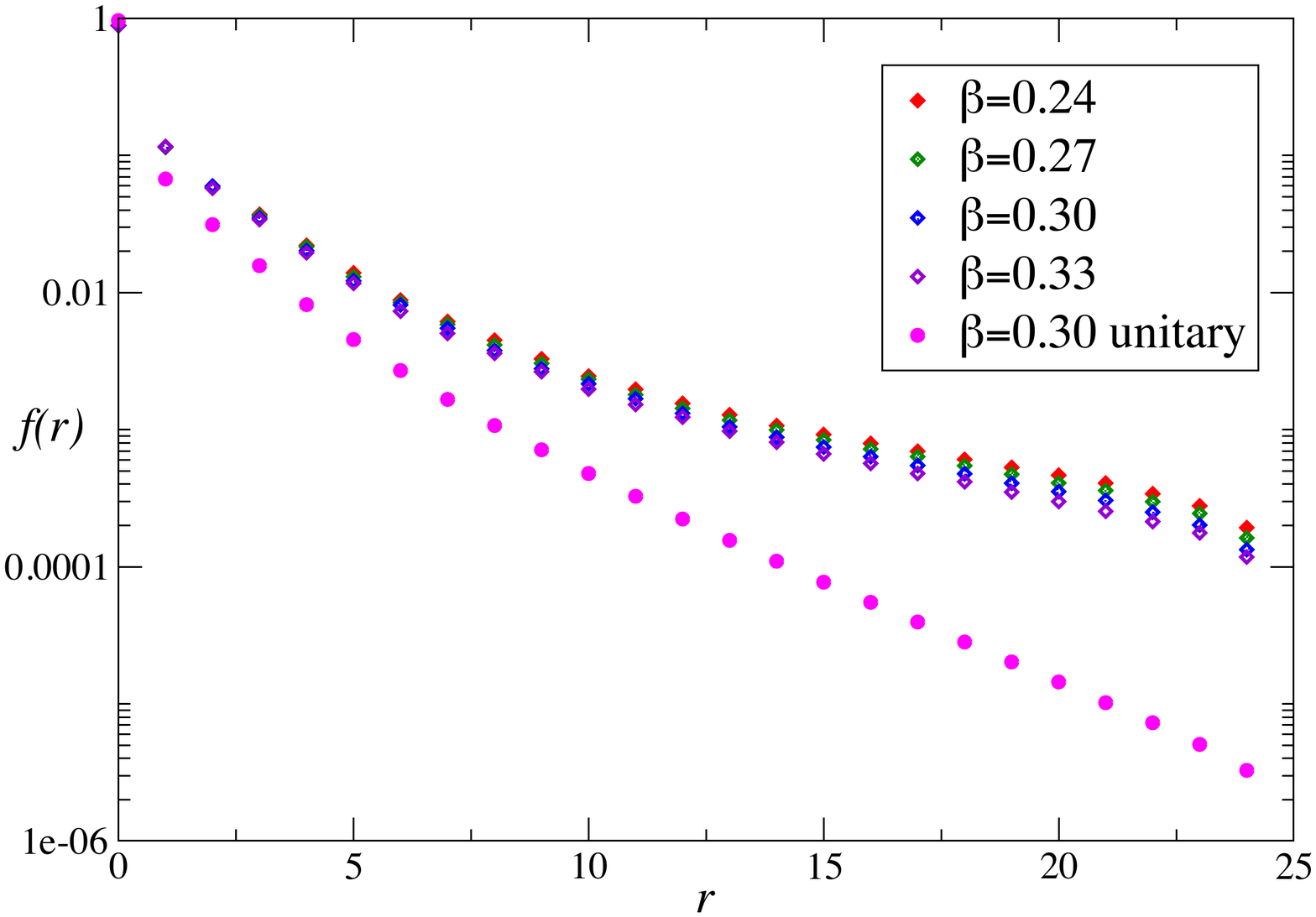}\\
  \caption{Localisation of the overlap $D$ with Shamir kernel and $L_s=48$ 
averaged over 32 sources on each of 5
   configurations}
  \label{fig:M3mass}
  \includegraphics[width=0.85\columnwidth,scale=0.8]{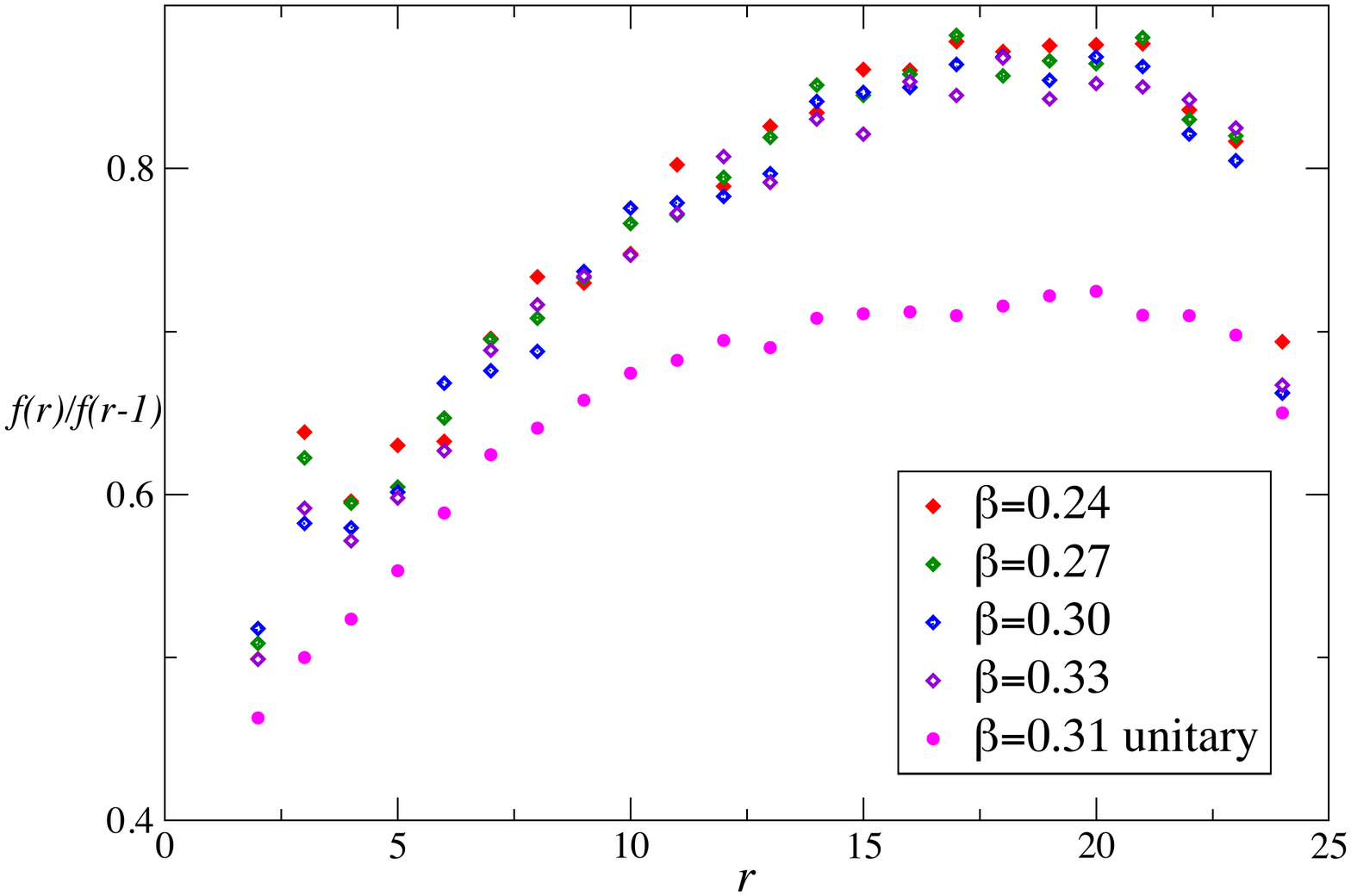}\\
  \caption{The decrement $f(r)/f(r-1)$ for the data of Fig.~\ref{fig:M3mass}}
  \label{fig:M3massd}
\end{center}
\end{figure}

The locality of $D_{OL}$ in the critical region is illustrated for the twisted
$im\bar\psi\gamma_3\psi$ mass term with $ma=0.005$ in
Figs.~\ref{fig:M3mass},\ref{fig:M3massd}.
Within the limitations imposed by a $16^3$ volume, Fig.~\ref{fig:M3mass}
is consistent with exponential falloff for $ra\gapprox10$. There is a mild
$\beta$-dependence; as expected the falloff slows down as the coupling gets
stronger. Also shown for comparison are data obtained with unitary link fields
$e^{\pm iA_\mu}$,
where exponential localisation is more manifest. 
Convergence to a meaningful
value of the decay rate on this small volume is seen to be difficult in
Fig.~\ref{fig:M3massd} 
showing the decrement from one $r$-value to the next, by plotting $f(r)/f(r-1)$.
As the coupling strength gets closer to criticality there is only marginal
variation, indicating locality is not jeopardised at a critical point.
Varying $m$ does not change the conclusions, nor does using the Wilson
kernel rather than Shamir.

\begin{figure}[tbp]
\begin{center}
  \includegraphics[width=0.85\columnwidth,scale=0.8]{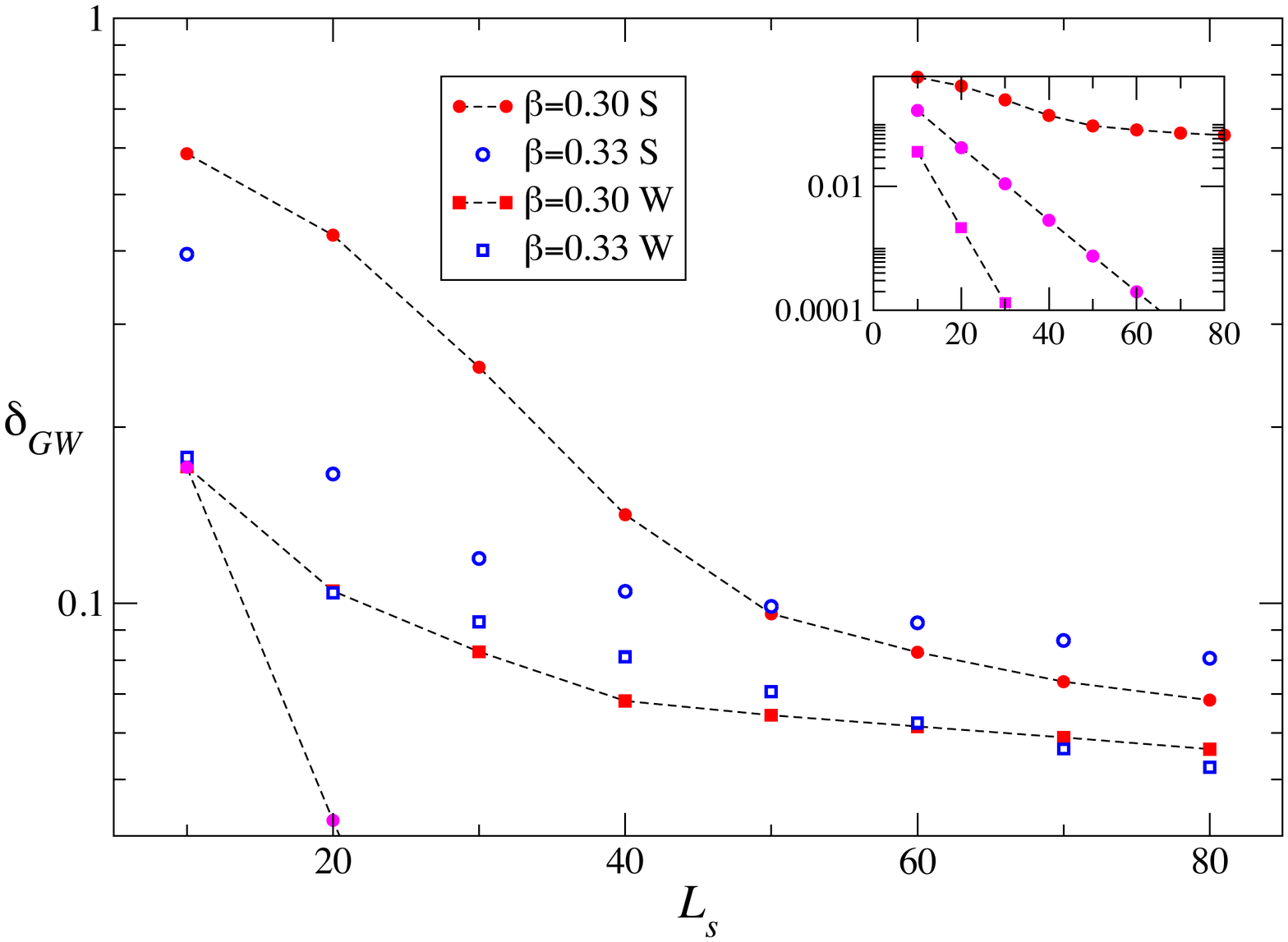}\\
\caption{GW error $\delta_{GW}$ calculated on a single configuration for
Shamir (S) and Wilson (W) kernels. The inset compares results obtained
using unitary link fields at $\beta=0.3$.}
  \label{fig:GWerror}
\end{center}
\end{figure}

We also examine the truncation/finite-$L_s$ error of the overlap/DWF  operator
via the GW term, as a means to assess recovery of U(2) symmetry. In the
$n\to\infty$ ($L_s\to\infty$) limits the GW error $\delta_{GW}$, given as
\begin{equation} 
\delta_{GW}=||(\gamma_3 D + D\gamma_3 - 2 D
\gamma_3 D)\phi||_{\infty} 
\label{eq:GWerror}
\end{equation} 
with $\phi$ a 
complex field chosen at random at each lattice site such that each component is
distributed uniformly in $(-1,1)$, should be exactly zero for zero mass. We use a single auxiliary field
configuration to
plot $\delta_{GW}$ in Fig.~\ref{fig:GWerror}. Although the boson field 
is generated  with a non-zero mass, the GW error is measured with $D(m=0)$.

The GW error vanishes only very slowly, 
and in fact $\delta_{GW}$ 
is numerically very similar to $\delta_h$  defined in
(\ref{eq:deltah}) (plotted as a function of $L_s$ in Fig.~12 of
\cite{Hands:2018vrd}). It is interesting that convergence with
the Wilson kernel is slightly faster as compared to the Shamir formulation, which
is surprising since with non-unitary links the Wilson kernel is not bounded
which should prejudice convergence~\cite{Kennedy:2006ax,Hands:2015dyp}.    
Again, for comparison results obtained with unitary link fields are included in
the inset;
here exponential falloff of the error is much sharper.
These results suggest 
that the very large values of $L_s$
needed for U($2N$) recovery in the critical region have their origin in the
non-unitary nature of the link fields in this formalism.



\section{Discussion}
\label{sec:discussion}
The main conclusion of this study is that we can state with much more confidence
that there is a phase transition associated with bilinear condensation 
in the Thirring model defined with $N=1$ domain
wall fermions, and hence that $N_c>1$ as originally suggested in
\cite{Hands:2018vrd}. We have also made the first steps towards characterising
the critical properties. The enhanced dataset we have generated demonstrates both
the importance and the stability of the $L_s\to\infty$ extrapolation, enabling 
fits to an RG-inspired equation of state with $\beta_c\simeq0.279(1)$ which work extremely well for
$\beta\gapprox\beta_c$. The fitted critical exponents are
significantly different from their mean-field values,  although
both Fig.~\ref{fig:decay_constants_16} and
Figs.~\ref{fig:betac_Ls},\ref{fig:exponents_Ls} suggest simulations with larger
$L_s$, perhaps even $L_s\sim O(10^2)$, will be needed in order to fully control the
predictions. Further encouraging signs are a susceptibility peak
of the expected shape seen in Fig.~\ref{fig:susc_light} and the $m$-dependence
of the bosonic action in the subcritical region seen in Fig.~\ref{fig:bosepeek}.

The result $N_c\gapprox1$ is in clear contast with predictions obtained with staggered
fermions, for the reasons reviewed in the Introduction, but more interestingly
also with $N_c=0.80(4)$ obtained with SLAC
fermions~\cite{Lenz:2019qwu}. We speculate that the two lattice approaches
describe different continuum theories, and that the bulk DWF formulation
followed here more closely conforms to a picture of the strong dynamics in which
the auxiliary boson $A_\mu$ resembles a gauge field. Ultimately, physics at a
QCP is  specified not by a Lagrangian density, either continuum-like or
regularised, but by more primitive considerations such as dimensionality, field
content, and of course the pattern of global symmetry breaking.
Interestingly, a recent study using the Conformal Bootstrap predicts 
$1<N_c<2$ for QED$_3$~\cite{Li:2018lyb}.

For the first time in Sec.~\ref{sec:locality} we have studied 
properties of the $2+1d$ overlap operator $D_{OL}$ associated with DWF in the
vicinity of the critical point. Fig.~\ref{fig:M3mass} suggests there exists 
a limit where $aD_{OL}\to0$, a necessary condition for the existence of a
continuum limit with conventional U($2N$) symmetry, although confirmation will
require studies on larger spacetime volumes. However, symmetry
recovery also requires the restoration of the Ginsparg-Wilson relations
(\ref{eq:GW}). Both the direct estimates shown in Fig.~\ref{fig:GWerror}
and the indirect measure via the residual $\delta_h$ shown in
Fig.~\ref{fig:deltapeek} suggest this is at best restored only very slowly in the
critical region. Note that $\delta_h\to0$ is not guaranteed even once
$O(e^{-\Delta L_s})$ corrections are applied to the order parameter
as described in Sec.~\ref{sec:Lsextrap}. It is now becoming apparent that this behaviour has its
origins in the use of non-unitary link fields in the Wilson operator $D_W$;
recall this originates in the desire to have only four-point fermion interactions
left once the auxiliary is integrated over. Fig.~\ref{fig:GWerror} is also a
motivation to explore measurement using $D_{OL}$ defined with the alternative
Wilson kernel, and/or 
with expansion order $n$ greater than the $L_s$ used in ensemble
generation with DWF.

It's also important to review areas where the simulation falls short of the QCP
ideal. As outlined in Sec.~\ref{sec:Ls48}, at fixed $L_s=48$ EoS fits fail to
describe the data with $\beta<\beta_c$, and the susceptibility plot
Fig.~\ref{fig:suscpeek} shows an inverted $m$-hierarchy when compared with the
expectation of Fig.~\ref{fig:model_susc}, and moreover requires a
seemingly arbitrary rescaling even to fit a single $m$ value. The 
behaviour at weaker couplings can be somewhat accommodated by replacing the LHS
of the EoS (\ref{eq:eos}) by a factor $m(m/m_0)^\alpha$. The inset of
Fig.~\ref{fig:model_susc} shows the susceptibility curves thus obtained with 
$m_0a=0.005$, $\alpha=0.3$. This fails to fix the hierarchy in the
critical region, however; it seems safer to conclude that $\chi_\ell$ 
defined in (\ref{eq:chi}) is {\em not\/} the second derivative of the free energy of a
2+1$d$ theory. A modification of the physical field prescription
(\ref{eq:physical}) may be needed, to allow for the possibility that the relevant
modes close to the walls ``leak'' somewhat into the bulk as $ma\to0$.
A related puzzle, again a failure to reconcile the observed behaviour of order parameter and
susceptibility data with possibly the same cause, is the 
breakdown of the axial Ward Identity noted in \cite{Hands:2016foa}.

The flattening of the order parameter at small $\beta$ seen in
Figs.~\ref{fig:eosfit_sc},\ref{fig:cond_light} remains a worrying aspect. 
Possibly the $L_s\to\infty$ extrapolation is not yet under control in the
broken region -- this scenario is supported by the results of pilot simulations
on $16^3\times64$ with $\beta=0.3$, $ma=0.005$, plotted in
Figs.~\ref{fig:cond_light},\ref{fig:susc_light} (the $L_s=64$ point lying on
the fitted curve in Fig.~\ref{fig:cond_light} should be regarded at this stage as
a coincidence).
Another possibility is that the scaling window where (\ref{eq:eos}) is applicable
is simply very narrow on the horizontal scale used in these figures. As things
stand, however, we have not yet demonstrated a clear separation between
$\beta_c$ and a putative $\beta^*$ where lattice artifacts dominate, and hence
do not yet have an understanding of the broken phase comparable with that for
$\beta>\beta_c$. The slow convergence to the U($2N$) limit discussed above may
prove a formidable obstacle; there is much work still to be done.

\section*{Acknowledgements}
This work was performed using the Cambridge Service for Data Driven Discovery
(CSD3), part of which is operated by the University of Cambridge Research
Computing on behalf of the STFC DiRAC HPC Facility (www.dirac.ac.uk). The DiRAC
component of CSD3 was funded by BEIS capital funding via STFC capital grants
ST/P002307/1 and ST/R002452/1 and STFC operations grant ST/R00689X/1. DiRAC is
part of the National e-Infrastructure. Additional work utilised the 
{\em Sunbird\/}  facility of Supercomputing Wales.
The work of SJH was supported 
by STFC grant ST/L000369/1, 
of MM in part by the Supercomputing Wales project, which is
part-funded by the European Regional Development Fund (ERDF) via Welsh
Government,
and of JW by an EPSRC studentship.

\end{document}